\documentclass[a4paper,10pt]{article} %PP
\usepackage{a4wide} %PP
\usepackage{graphicx}
\usepackage{amssymb}

\newcommand{\tr}{\mathrel{\rm tr}}
\newcommand{\la}{\lesssim}
\newcommand{\ga}{\gtrsim}
\newcommand{\farcs}{\hbox{$.\!\!^{\prime\prime}$}}

\begin{document}

\title{Lectures on Gravitational Lensing} %PP

%JO \author[R. Narayan \& M. Bartelmann]
\author %PP
	{Ramesh NARAYAN \\
	Harvard-Smithsonian Center for Astrophysics \\
	60 Garden Street \\
	Cambridge, MA 02138, USA \\
	\and
	Matthias BARTELMANN \\
	Max-Planck-Institut f\"ur Astrophysik \\
	P.O. Box 1523, D--85740 Garching, Germany}

\maketitle %PP

\begin{abstract}
These lectures give an introduction to Gravitational Lensing. We
discuss lensing by point masses, lensing by galaxies, and lensing by
clusters and larger-scale structures in the Universe. The relevant
theory is developed and applications to astrophysical problems are
discussed.
\end{abstract}

\setcounter{tocdepth}{2}\clearpage\tableofcontents\clearpage %PP

%JO \chapter{Lectures on Gravitational Lensing}

\section{Introduction}
\label{NB:sect1}

One of the consequences of Einstein's General Theory of Relativity is
that light rays are deflected by gravity. Although this discovery was
made only in this century, the possibility that there could be such a
deflection had been suspected much earlier, by Newton and Laplace
among others. Soldner (1804) calculated the magnitude of the
deflection due to the Sun, assuming that light consists of material
particles and using Newtonian gravity. Later, Einstein (1911) employed
the equivalence principle to calculate the deflection angle and
re-derived Soldner's formula. Later yet, Einstein (1915) applied the
full field equations of General Relativity and discovered that the
deflection angle is actually twice his previous result, the factor of
two arising because of the curvature of the metric. According to this
formula, a light ray which tangentially grazes the surface of the Sun
is deflected by $1\farcs7$. Einstein's final result was confirmed in
1919 when the apparent angular shift of stars close to the limb of the
Sun (see Fig.\ \ref{NB:fig101}) was measured during a total solar
eclipse (Dyson, Eddington, \& Davidson 1920). The quantitative
agreement between the measured shift and Einstein's prediction was
immediately perceived as compelling evidence in support of the theory
of General Relativity. The deflection of light by massive bodies, and
the phenomena resulting therefrom, are now referred to as {\em
Gravitational Lensing\/}.

\begin{figure}[ht] %PP
%JO \begin{figure}
\begin{center}
\begin{minipage}[t]{0.9\hsize}
\end{minipage}
\end{center}
\caption{Angular deflection of a ray of light passing close to the
  limb of the Sun. Since the light ray is bent toward the Sun, the
  apparent positions of stars move away from the Sun.}
\label{NB:fig101}
\end{figure}

Eddington (1920) noted that under certain conditions there may be
multiple light paths connecting a source and an observer. This implies
that gravitational lensing can give rise to multiple images of a
single source. Chwolson (1924) considered the creation of fictitious
double stars by gravitational lensing of stars by stars, but did not
comment on whether the phenomenon could actually be observed.
Einstein (1936) discussed the same problem and concluded that there is
little chance of observing lensing phenomena caused by stellar-mass
lenses. His reason was that the angular image splitting caused by a
stellar-mass lens is too small to be resolved by an optical telescope.

Zwicky (1937a) elevated gravitational lensing from a curiosity to a
field with great potential when he pointed out that galaxies can split
images of background sources by a large enough angle to be
observed. At that time, galaxies were commonly believed to have masses
of $\sim10^9\,M_\odot$. However, Zwicky had applied the virial theorem
to the Virgo and Coma clusters of galaxies and had derived galaxy
masses of $\sim4\times10^{11}\,M_\odot$. Zwicky argued that the
deflection of light by galaxies would not only furnish an additional
test of General Relativity, but would also magnify distant galaxies
which would otherwise remain undetected, and would allow accurate
determination of galaxy masses. Zwicky (1937b) even calculated the
probability of lensing by galaxies and concluded that it is on the
order of one per cent for a source at reasonably large redshift.

Virtually all of Zwicky's predictions have come true. Lensing by
galaxies is a major sub-discipline of gravitational lensing today. The
most accurate mass determinations of the central regions of galaxies
are due to gravitational lensing, and the cosmic telescope effect of
gravitational lenses has enabled us to study faint and distant
galaxies which happen to be strongly magnified by galaxy clusters. The
statistics of gravitational lensing events, whose order of magnitude
Zwicky correctly estimated, offers one of the promising ways of
inferring cosmological parameters.

In a stimulating paper, Refsdal (1964) described how the Hubble
constant $H_0$ could in principle be measured through gravitational
lensing of a variable source. Since the light travel times for the
various images are unequal, intrinsic variations of the source would
be observed at different times in the images. The time delay between
images is proportional to the difference in the absolute lengths of
the light paths, which in turn is proportional to $H_0^{-1}$. Thus, if
the time delay is measured and if an accurate model of a lensed source
is developed, the Hubble constant could be measured.

All of these ideas on gravitational lensing remained mere speculation
until real examples of gravitational lensing were finally discovered.
The stage for this was set by the discovery of quasars (Schmidt 1963)
which revealed a class of sources that is ideal for studying the
effects of gravitational lensing. Quasars are distant, and so the
probability that they are lensed by intervening galaxies is
sufficiently large. Yet, they are bright enough to be detected even at
cosmological distances. Moreover, their optical emission region is
very compact, much smaller than the typical scales of galaxy
lenses. The resulting magnifications can therefore be very large, and
multiple image components are well separated and easily detected.

Walsh, Carswell, \& Weymann (1979) discovered the first example of
gravitational lensing, the quasar QSO 0957+561A,B. This source
consists of two images, A and B, separated by $6''$. Evidence that
0957+561A,B does indeed correspond to twin lensed images of a single
QSO is provided by (i) the similarity of the spectra of the two
images, (ii) the fact that the flux ratio between the images is
similar in the optical and radio wavebands, (iii) the presence of a
foreground galaxy between the images, and (iv) VLBI observations which
show detailed correspondence between various knots of emission in the
two radio images. Over a dozen convincing examples of multiple-imaged
quasars are known today (Keeton \& Kochanek 1996) and the list
continues to grow.

Paczy\'nski (1986b) revived the idea of lensing of stars by stars when
he showed that at any given time one in a million stars in the Large
Magellanic Cloud (LMC) might be measurably magnified by the
gravitational lens effect of an intervening star in the halo of our
Galaxy. The magnification events, which are called {\em
microlensing\/} events, have time scales between two hours and two
years for lens masses between $10^{-6}\,M_\odot$ and
$10^2\,M_\odot$. Initially, it was believed that the proposed
experiment of monitoring the light curves of a million stars would
never be feasible, especially since the light curves have to be
sampled frequently and need to be distinguished from light curves of
intrinsically variable stars.  Nevertheless, techniques have advanced
so rapidly that today four separate collaborations have successfully
detected microlensing events (Alcock et al.\ 1993; Aubourg et al.\
1993; Udalski et al.\ 1993; Alard 1995), and this field has developed
into an exciting method for studying the nature and distribution of
mass in our Galaxy.

{\em Einstein rings\/}, a particularly interesting manifestation of
gravitational lensing, were discovered first in the radio waveband by
Hewitt et al.\ (1987). About half a dozen radio rings are now known
and these sources permit the most detailed modeling yet of the mass
distributions of lensing galaxies.

Gravitational lensing by galaxy clusters had been considered
theoretically even before the discovery of QSO 0957+561. The subject
entered the observational realm with the discovery of giant blue
luminous {\em arcs\/} in the galaxy clusters A~370 and Cl~2244
(Soucail et al.\ 1987a,b; Lynds \& Petrosian 1986). Paczy\'nski (1987)
proposed that the arcs are the images of background galaxies which are
strongly distorted and elongated by the gravitational lens effect of
the foreground cluster. This explanation was confirmed when the first
arc redshifts were measured and found to be significantly greater than
that of the clusters.

Apart from the spectacular giant luminous arcs, which require special
alignment between the cluster and the background source, clusters also
coherently distort the images of other faint background galaxies
(Tyson 1988). These distortions are mostly weak, and the corresponding
images are referred to as {\em arclets\/} (Fort et al.\ 1988; Tyson,
Valdes, \& Wenk 1990). Observations of arclets can be used to
reconstruct parameter-free, two-dimensional mass maps of the lensing
cluster (Kaiser \& Squires 1993). This technique has attracted a great
deal of interest, and two-dimensional maps have been obtained of
several galaxy clusters (Bonnet et al.\ 1993; Bonnet, Mellier, \& Fort
1994; Fahlman et al.\ 1994; Broadhurst 1995; Smail et al.\ 1995; Tyson
\& Fischer 1995; Squires et al.\ 1996; Seitz et al.\ 1996).

As this brief summary indicates, gravitational lensing manifests
itself through a very broad and interesting range of phenomena. At the
same time, lensing has developed into a powerful tool to study a host
of important questions in astrophysics. The applications of
gravitational lensing may be broadly classified under three
categories:

\begin{itemize}

\item[--] The magnification effect enables us to observe objects which
are too distant or intrinsically too faint to be observed without
lensing. Lenses therefore act as ``cosmic telescopes'' and allow us to
infer source properties far below the resolution limit or sensitivity
limit of current observations. However, since we do not have the
ability to point this telescope at any particular object of
interest but have to work with whatever nature gives us, the results
have been only modestly interesting.

\item[--] Gravitational lensing depends solely on the projected,
two-dimensional mass distribution of the lens, and is independent of
the luminosity or composition of the lens. Lensing therefore offers an
ideal way to detect and study dark matter, and to explore the growth
and structure of mass condensations in the universe.

\item[--] Many properties of individual lens systems or samples of
lensed objects depend on the age, the scale, and the overall geometry
of the universe. The Hubble constant, the cosmological constant, and
the density parameter of the universe can be significantly constrained
through lensing.
\end{itemize}

The article is divided into three main sections. Sect.\ \ref{NB:sect2}
discusses the effects of point-mass lenses, Sect.\ \ref{NB:sect3}
considers galaxy-scale lenses, and Sect.\ \ref{NB:sect4} discusses
lensing by galaxy clusters and large-scale structure in the
universe. References to the original literature are given throughout
the text. The following are some general or specialized review
articles and monographs:

\begin{center}
{\bf Monograph}
\end{center}

\begin{itemize}

\item[--] Schneider, P., Ehlers, J., \& Falco, E.E. 1992, {\em
Gravitational Lenses\/} (Berlin: Springer Verlag)

\end{itemize}

\begin{center}
{\bf General Reviews}
\end{center}

\begin{itemize}

\item[--] Blandford, R.D., \& Narayan, R. 1992, {\em Cosmological
Applications of Gravitational Lensing\/}, Ann. Rev. Astr. Ap., 30, 311

\item[--] Refsdal, S., \& Surdej, J. 1994, {\em Gravitational
Lenses\/}, Rep. Progr. Phys., 57, 117

\item[--] Schneider, P. 1996, {\em Cosmological Applications of
Gravitational Lensing\/}, in: {\em The universe at high-$z$,
large-scale structure and the cosmic microwave background}, Lecture
Notes in Physics, eds. E. Mart{\'\i}nez-Gonz\'alez \& J.L. Sanz
(Berlin: Springer Verlag)

\item[--] Wu, X.-P. 1996, {\em Gravitational Lensing in the
Universe\/}, Fundamentals of Cosmic Physics, 17, 1

\end{itemize}

\begin{center}
{\bf Special Reviews}
\end{center}

\begin{itemize}

\item[--] Fort, B., \& Mellier, Y. 1994, {\em Arc(let)s in Clusters of
Galaxies\/}, Astr. Ap. Rev., 5, 239

\item[--] Bartelmann, M., \& Narayan, R. 1995, {\em Gravitational
Lensing and the Mass Distribution of Clusters\/}, in: {\em Dark
Matter\/}, AIP Conf. Proc. 336, eds. S.S. Holt \& C.L. Bennett (New
York: AIP Press)

\item[--] Keeton II, C.R. \& Kochanek, C.S. 1996, {\em Summary of Data
on Secure Multiply-Imaged Systems\/}, in: {\em Cosmological
Applications of Gravitational Lensing\/}, IAU Symp. 173,
eds. C.S. Kochanek \& J.N. Hewitt

\item[--] Paczy\'nski, B. 1996, {\em Gravitational Microlensing in the
Local Group\/}, Ann. Rev. Astr. Ap., 34, 419

\item[--] Roulet, E., \& Mollerach, S. 1997, {\em Microlensing\/},
Physics Reports, 279, 67

\end{itemize}

\section{Lensing by Point Masses in the Universe}
\label{NB:sect2}

\subsection{Basics of Gravitational Lensing}
\label{NB:subsect2.1}

The propagation of light in arbitrary curved spacetimes is in general
a complicated theoretical problem. However, for almost all cases of
relevance to gravitational lensing, we can assume that the overall
geometry of the universe is well described by the
Friedmann-Lema{\^\i}tre-Robertson-Walker metric and that the matter
inhomogeneities which cause the lensing are no more than local
perturbations. Light paths propagating from the source past the lens
to the observer can then be broken up into three distinct zones. In
the first zone, light travels from the source to a point close to the
lens through unperturbed spacetime. In the second zone, near the lens,
light is deflected. Finally, in the third zone, light again travels
through unperturbed spacetime. To study light deflection close to the
lens, we can assume a locally flat, Minkowskian spacetime which is
weakly perturbed by the Newtonian gravitational potential of the mass
distribution constituting the lens. This approach is legitimate if the
Newtonian potential $\Phi$ is small, $|\Phi|\ll c^2$, and if the
peculiar velocity $v$ of the lens is small, $v\ll c$.

These conditions are satisfied in virtually all cases of astrophysical
interest. Consider for instance a galaxy cluster at redshift $\sim0.3$
which deflects light from a source at redshift $\sim1$. The distances
from the source to the lens and from the lens to the observer are
$\sim1$~Gpc, or about three orders of magnitude larger than the
diameter of the cluster. Thus zone 2 is limited to a small local
segment of the total light path. The relative peculiar velocities in a
galaxy cluster are $\sim10^3$ km s$^{-1}\ll c$, and the Newtonian
potential is $|\Phi|<10^{-4}\,c^2\ll c^2$, in agreement with the
conditions stated above.

\subsubsection{Effective Refractive Index of a Gravitational Field}
\label{NB:subsubsect2.1.1}

In view of the simplifications just discussed, we can describe light
propagation close to gravitational lenses in a locally Minkowskian
spacetime perturbed by the gravitational potential of the lens to
first post-Newtonian order. The effect of spacetime curvature on the
light paths can then be expressed in terms of an effective index of
refraction $n$, which is given by (e.g.\ Schneider et al.\ 1992)
\begin{equation}
  n = 1 - \frac{2}{c^2}\,\Phi = 1 + \frac{2}{c^2}\,|\Phi|\;.
\label{eq:2.1}
\end{equation}
Note that the Newtonian potential is negative if it is defined such
that it approaches zero at infinity. As in normal geometrical optics,
a refractive index $n>1$ implies that light travels slower than in
free vacuum. Thus, the effective speed of a ray of light in a
gravitational field is
\begin{equation}
  v = \frac{c}{n} \simeq c - \frac{2}{c}\,|\Phi|\;.
\label{eq:2.2}
\end{equation}

Figure \ref{NB:fig2.1} shows the deflection of light by a glass prism.
The speed of light is reduced inside the prism. This reduction of
speed causes a delay in the arrival time of a signal through the prism
relative to another signal traveling at speed $c$. In addition, it
causes wavefronts to tilt as light propagates from one medium to
another, leading to a bending of the light ray around the thick end of
the prism.

\begin{figure}[ht] %PP
%JO \begin{figure}
\begin{center}
\begin{minipage}[t]{0.6\hsize}
\end{minipage}
\end{center}
\caption{Light deflection by a prism. The refractive index $n>1$ of
  the glass in the prism reduces the effective speed of light to
  $c/n$. This causes light rays to be bent around the thick end of the
  prism, as indicated. The dashed lines are wavefronts. Although the
  geometrical distance between the wavefronts along the two rays is
  different, the travel time is the same because the ray on the left
  travels through a larger thickness of glass.}
\label{NB:fig2.1}
\end{figure}

The same effects are seen in gravitational lensing. Because the
effective speed of light is reduced in a gravitational field, light
rays are delayed relative to propagation in vacuum. The total time
delay $\Delta t$ is obtained by integrating over the light path from
the observer to the source:
\begin{equation}
  \Delta t = \int_{\rm source}^{\rm observer}\,
  \frac{2}{c^3}\,|\Phi|\,dl\;.
\label{eq:2.4}
\end{equation}
This is called the Shapiro delay (Shapiro 1964).

As in the case of the prism, light rays are deflected when they pass
through a gravitational field. The deflection is the integral along
the light path of the gradient of $n$ perpendicular to the light path,
i.e.
\begin{equation}
  \vec{\hat\alpha} = -\int\vec\nabla_\perp n\,dl
  = \frac{2}{c^2}\,\int\vec\nabla_\perp\Phi\,dl\;.
\label{eq:2.5}
\end{equation}
In all cases of interest the deflection angle is very small. We can
therefore simplify the computation of the deflection angle
considerably if we integrate $\vec\nabla_\perp n$ not along the
deflected ray, but along an unperturbed light ray with the same impact
parameter. (As an aside we note that while the procedure is
straightforward with a single lens, some care is needed in the case of
multiple lenses at different distances from the source. With multiple
lenses, one takes the unperturbed ray from the source as the reference
trajectory for calculating the deflection by the first lens, the
deflected ray from the first lens as the reference unperturbed ray for
calculating the deflection by the second lens, and so on.)

\begin{figure}[ht] %PP
%JO \begin{figure}
\begin{center}
\begin{minipage}[t]{0.6\hsize}
\end{minipage}
\end{center}
\caption{Light deflection by a point mass. The unperturbed ray passes
  the mass at impact parameter $b$ and is deflected by the angle
  $\hat\alpha$. Most of the deflection occurs within $\Delta z\sim\pm
  b$ of the point of closest approach.}
\label{NB:fig2.2}
\end{figure}

As an example, we now evaluate the deflection angle of a point mass
$M$ (cf.\ Fig.\ \ref{NB:fig2.2}). The Newtonian potential of the lens
is
\begin{equation}
  \Phi(b,z) = -\frac{GM}{(b^2+z^2)^{1/2}}\;,
\label{eq:2.6}
\end{equation}
where $b$ is the impact parameter of the unperturbed light ray, and
$z$ indicates distance along the unperturbed light ray from the point
of closest approach. We therefore have
\begin{equation}
  \vec\nabla_\perp\Phi(b,z) = \frac{GM\,\vec b}{(b^2+z^2)^{3/2}}\;,
\label{eq:2.7}
\end{equation}
where $\vec b$ is orthogonal to the unperturbed ray and points toward
the point mass. Equation (\ref{eq:2.7}) then yields the deflection
angle
\begin{equation}
  \hat\alpha = \frac{2}{c^2}\,
  \int\vec\nabla_\perp\Phi\,dz
  = \frac{4GM}{c^2b}\;.
\label{eq:2.8}
\end{equation}
Note that the Schwarzschild radius of a point mass is
\begin{equation}
  R_{\rm S} = \frac{2GM}{c^2}\;,
\label{eq:2.9}
\end{equation}
so that the deflection angle is simply twice the inverse of the impact
parameter in units of the Schwarzschild radius. As an example, the
Schwarzschild radius of the Sun is $2.95$~km, and the solar radius is
$6.96\times10^5$~km. A light ray grazing the limb of the Sun is
therefore deflected by an angle
$(5.9/7.0)\times10^{-5}\,\hbox{radians}=1\farcs7$.

\subsubsection{Thin Screen Approximation}
\label{NB:subsubsect2.1.2}

Figure \ref{NB:fig2.2} illustrates that most of the light deflection
occurs within $\Delta z\sim\pm b$ of the point of closest encounter
between the light ray and the point mass. This $\Delta z$ is typically
much smaller than the distances between observer and lens and between
lens and source. The lens can therefore be considered thin compared to
the total extent of the light path. The mass distribution of the lens
can then be projected along the line-of-sight and be replaced by a
mass sheet orthogonal to the line-of-sight. The plane of the mass
sheet is commonly called the lens plane. The mass sheet is
characterized by its surface mass density
\begin{equation}
  \Sigma(\vec\xi) = \int\,\rho(\vec\xi,z)\,dz\;,
\label{eq:2.10}
\end{equation}
where $\vec\xi$ is a two-dimensional vector in the lens plane. The
deflection angle at position $\vec\xi$ is the sum of the deflections
due to all the mass elements in the plane:
\begin{equation}
  \vec{\hat\alpha}(\vec\xi) = \frac{4G}{c^2}\,\int\,
  \frac{(\vec\xi-\vec\xi')\Sigma(\vec\xi')}{|\vec\xi-\vec\xi'|^2}
  \,d^2\xi'\;.
\label{eq:2.11}
\end{equation}
Figure \ref{NB:fig2.3} illustrates the situation.

\begin{figure}[ht] %PP
%JO \begin{figure}
\begin{center}
\begin{minipage}[t]{0.6\hsize}
\end{minipage}
\end{center}
\caption{A light ray which intersects the lens plane at $\vec\xi$ is
  deflected by an angle $\vec{\hat\alpha}(\vec\xi)$.}
\label{NB:fig2.3}
\end{figure}

In general, the deflection angle is a two-component vector. In the
special case of a circularly symmetric lens, we can shift the
coordinate origin to the center of symmetry and reduce light
deflection to a one-dimensional problem. The deflection angle then
points toward the center of symmetry, and its modulus is
\begin{equation}
  \hat\alpha(\xi) = \frac{4GM(\xi)}{c^2\xi}\;,
\label{eq:2.12}
\end{equation}
where $\xi$ is the distance from the lens center and $M(\xi)$ is the
mass enclosed within radius $\xi$,
\begin{equation}
  M(\xi) = 2\pi\,\int_0^\xi\,\Sigma(\xi')\xi'\,d\xi'\;.
\label{eq:2.13}
\end{equation}

\subsubsection{Lensing Geometry and Lens Equation}
\label{NB:subsubsect2.1.3}

The geometry of a typical gravitational lens system is shown in Fig.\
\ref{NB:fig2.4}. A light ray from a source S is deflected by the angle
$\vec{\hat\alpha}$ at the lens and reaches an observer O. The angle
between the (arbitrarily chosen) optic axis and the true source
position is $\vec\beta$, and the angle between the optic axis and the
image I is $\vec\theta$. The (angular diameter) distances between
observer and lens, lens and source, and observer and source are
$D_{\rm d}$, $D_{\rm ds}$, and $D_{\rm s}$, respectively.

\begin{figure}[ht] %PP
%JO \begin{figure}
\begin{center}
\begin{minipage}[t]{0.6\hsize}
\end{minipage}
\end{center}
\caption{Illustration of a gravitational lens system. The light ray
  propagates from the source S at transverse distance $\eta$ from the
  optic axis to the observer O, passing the lens at transverse
  distance $\xi$. It is deflected by an angle $\hat\alpha$. The
  angular separations of the source and the image from the optic axis
  as seen by the observer are $\beta$ and $\theta$, respectively. The
  reduced deflection angle $\alpha$ and the actual deflection angle
  $\hat\alpha$ are related by eq.\ (\protect\ref{eq:2.14}). The
  distances between the observer and the source, the observer and the
  lens, and the lens and the source are $D_{\rm s}$, $D_{\rm d}$, and
  $D_{\rm ds}$, respectively.}
\label{NB:fig2.4}
\end{figure}

It is now convenient to introduce the reduced deflection angle
\begin{equation}
  \vec\alpha = \frac{D_{\rm ds}}{D_{\rm s}}\,\vec{\hat\alpha}\;.
\label{eq:2.14}
\end{equation}
From Fig.\ \ref{NB:fig2.4} we see that $\theta D_{\rm s}=\beta D_{\rm
s}-{\hat\alpha}D_{\rm ds}$. Therefore, the positions of the source and
the image are related through the simple equation
\begin{equation}
  \vec\beta = \vec\theta - \vec\alpha(\vec\theta)\;.
\label{eq:2.15}
\end{equation}
Equation (\ref{eq:2.15}) is called the {\em lens equation\/}, or
ray-tracing equation. It is nonlinear in the general case, and so it
is possible to have multiple images $\vec\theta$ corresponding to a
single source position $\vec\beta$. As Fig.\ \ref{NB:fig2.4} shows,
the lens equation is trivial to derive and requires merely that the
following Euclidean relation should exist between the angle enclosed
by two lines and their separation,
\begin{equation}
  {\rm separation} = {\rm angle}\times{\rm distance}\;.
\label{eq:2.16}
\end{equation}
It is not obvious that the same relation should also hold in curved
spacetimes. However, if the distances $D_{\rm d,s,ds}$ are {\em
defined\/} such that eq.\ (\ref{eq:2.16}) holds, then the lens
equation must obviously be true. Distances so defined are called
angular-diameter distances, and eqs.\ (\ref{eq:2.14}), (\ref{eq:2.15})
are valid only when these distances are used. Note that in general
$D_{\rm ds}\ne D_{\rm s}-D_{\rm d}$.

As an instructive special case consider a lens with a constant
surface-mass density. From eq.\ (\ref{eq:2.12}), the (reduced)
deflection angle is
\begin{equation}
  \alpha(\theta) =
  \frac{D_{\rm ds}}{D_{\rm s}}\,\frac{4G}{c^2\xi}\,(\Sigma\pi\xi^2) =
  \frac{4\pi G\,\Sigma}{c^2}\,
  \frac{D_{\rm d}D_{\rm ds}}{D_{\rm s}}\,\theta\;,
\label{eq:2.17}
\end{equation}
where we have set $\xi=D_{\rm d}\theta$. In this case, the lens
equation is linear; that is, $\beta\propto\theta$. Let us define a
critical surface-mass density
\begin{equation}
  \Sigma_{\rm cr} =
  \frac{c^2}{4\pi G}\,\frac{D_{\rm s}}{D_{\rm d}D_{\rm ds}} =
  0.35\,{\rm g}\,{\rm cm}^{-2}\,
  \left(\frac{D}{1\,{\rm Gpc}}\right)^{-1}\;,
\label{eq:2.18}
\end{equation}
where the effective distance $D$ is defined as the combination of
distances
\begin{equation}
  D = \frac{D_{\rm d}D_{\rm ds}}{D_{\rm s}}\;.
\label{eq:2.19}
\end{equation}
For a lens with a constant surface mass density $\Sigma_{\rm cr}$, the
deflection angle is $\alpha(\theta)=\theta$, and so $\beta=0$ for all
$\theta$. Such a lens focuses perfectly, with a well-defined focal
length. A typical gravitational lens, however, behaves quite
differently. Light rays which pass the lens at different impact
parameters cross the optic axis at different distances behind the
lens. Considered as an optical device, a gravitational lens therefore
has almost all aberrations one can think of. However, it does not have
any chromatic aberration because the geometry of light paths is
independent of wavelength.

A lens which has $\Sigma>\Sigma_{\rm cr}$ somewhere within it is
referred to as being {\em supercritical\/}. Usually, multiple imaging
occurs only if the lens is supercritical, but there are exceptions to
this rule (e.g., Subramanian \& Cowling 1986).

\subsubsection{Einstein Radius}
\label{NB:subsubsect2.1.4}

Consider now a circularly symmetric lens with an arbitrary mass
profile. According to eqs.\ (\ref{eq:2.12}) and (\ref{eq:2.14}), the
lens equation reads
\begin{equation}
  \beta(\theta) = \theta-\frac{D_{\rm ds}}{D_{\rm d}D_{\rm s}}\,
  \frac{4GM(\theta)}{c^2\,\theta}\;.
\label{eq:2.20}
\end{equation}
Due to the rotational symmetry of the lens system, a source which lies
exactly on the optic axis ($\beta=0$) is imaged as a ring if the lens
is supercritical. Setting $\beta=0$ in eq.\ (\ref{eq:2.20}) we obtain
the radius of the ring to be
\begin{equation}
  \theta_{\rm E} = \left[\frac{4GM(\theta_{\rm E})}{c^2}\,
  \frac{D_{\rm ds}}{D_{\rm d}D_{\rm s}}\right]^{1/2}\;.
\label{eq:2.21}
\end{equation}
This is referred to as the {\em Einstein radius\/}. Figure
\ref{NB:fig2.5} illustrates the situation. Note that the Einstein
radius is not just a property of the lens, but depends also on the
various distances in the problem.

\begin{figure}[ht] %PP
%JO \begin{figure}
\begin{center}
\begin{minipage}[t]{0.6\hsize}
\end{minipage}
\end{center}
\caption{A source S on the optic axis of a circularly symmetric lens
  is imaged as a ring with an angular radius given by the Einstein
  radius $\theta_{\rm E}$.}
\label{NB:fig2.5}
\end{figure}

The Einstein radius provides a natural angular scale to describe the
lensing geometry for several reasons. In the case of multiple imaging,
the typical angular separation of images is of order $2\theta_{\rm
E}$. Further, sources which are closer than about $\theta_{\rm E}$ to
the optic axis experience strong lensing in the sense that they are
significantly magnified, whereas sources which are located well
outside the Einstein ring are magnified very little. In many lens
models, the Einstein ring also represents roughly the boundary between
source positions that are multiply-imaged and those that are only
singly-imaged. Finally, by comparing eqs.\ (\ref{eq:2.18}) and
(\ref{eq:2.21}) we see that the mean surface mass density inside the
Einstein radius is just the critical density $\Sigma_{\rm cr}$.

For a point mass $M$, the Einstein radius is given by
\begin{equation}
  \theta_{\rm E} = \left(\frac{4GM}{c^2}\,
  \frac{D_{\rm ds}}{D_{\rm d}D_{\rm s}}\right)^{1/2}\;.
\label{eq:2.22}
\end{equation}
To give two illustrative examples, we consider lensing by a star in
the Galaxy, for which $M\sim M_\odot$ and $D\sim10$~kpc, and lensing
by a galaxy at a cosmological distance with $M\sim10^{11}\,M_\odot$
and $D\sim1$~Gpc. The corresponding Einstein radii are
\begin{eqnarray}
  \theta_{\rm E} &=&
  (0.9\,{\rm mas})\,\left(\frac{M}{M_\odot}\right)^{1/2}\,
  \left(\frac{D}{10\,{\rm kpc}}\right)^{-1/2}\;,\nonumber\\
  \theta_{\rm E} &=&
  (0\farcs9)\,\left(\frac{M}{10^{11}\,M_\odot}\right)^{1/2}\,
  \left(\frac{D}{{\rm Gpc}}\right)^{-1/2}\;.\nonumber\\
\label{eq:2.23}
\end{eqnarray}

\subsubsection{Imaging by a Point Mass Lens}
\label{NB:subsubsect2.1.5}

For a point mass lens, we can use the Einstein radius (\ref{eq:2.21})
to rewrite the lens equation in the form
\begin{equation}
  \beta = \theta - \frac{\theta_{\rm E}^2}{\theta}\;.
\label{eq:2.24}
\end{equation}
This equation has two solutions,
\begin{equation}
  \theta_\pm = \frac{1}{2}\,\left(
  \beta\pm\sqrt{\beta^2+4\theta_{\rm E}^2}\right)\;.
\label{eq:2.25}
\end{equation}
Any source is imaged twice by a point mass lens. The two images are on
either side of the source, with one image inside the Einstein ring and
the other outside. As the source moves away from the lens (i.e. as
$\beta$ increases), one of the images approaches the lens and becomes
very faint, while the other image approaches closer and closer to the
true position of the source and tends toward a magnification of unity.

\begin{figure}[ht] %PP
%JO \begin{figure}
\begin{center}
\begin{minipage}[t]{0.6\hsize}
\end{minipage}
\end{center}
\caption{Relative locations of the source S and images I$_+$, I$_-$
  lensed by a point mass M. The dashed circle is the Einstein ring
  with radius $\theta_{\rm E}$. One image is inside the Einstein ring
  and the other outside.}
\label{NB:fig2.6}
\end{figure}

Gravitational light deflection preserves surface brightness (because
of Liouville's theorem), but gravitational lensing changes the
apparent solid angle of a source. The total flux received from a
gravitationally lensed image of a source is therefore changed in
proportion to the ratio between the solid angles of the image and the
source,
\begin{equation}
  \hbox{magnification} =
  \frac{\hbox{image area}}{\hbox{source area}}\;.
\label{eq:2.26}
\end{equation}
Figure \ref{NB:fig2.7} shows the magnified images of a source lensed
by a point mass.

\begin{figure}[ht] %PP
%JO \begin{figure}
\begin{center}
\begin{minipage}[t]{0.6\hsize}
\end{minipage}
\end{center}
\caption{Magnified images of a source lensed by a point mass.}
\label{NB:fig2.7}
\end{figure}

For a circularly symmetric lens, the magnification factor $\mu$ is
given by
\begin{equation}
  \mu = \frac{\theta}{\beta}\,\frac{d\theta}{d\beta}\;.
\label{eq:2.27}
\end{equation}
For a point mass lens, which is a special case of a circularly
symmetric lens, we can substitute for $\beta$ using the lens equation
(\ref{eq:2.24}) to obtain the magnifications of the two images,
\begin{equation}
  \mu_\pm = \left[
    1-\left(\frac{\theta_{\rm E}}{\theta_\pm}\right)^4
  \right]^{-1} =
  \frac{u^2+2}{2u\sqrt{u^2+4}}\pm\frac{1}{2}\;,
\label{eq:2.28}
\end{equation}
where $u$ is the angular separation of the source from the point mass
in units of the Einstein angle, $u=\beta\theta_{\rm E}^{-1}$. Since
$\theta_-<\theta_{\rm E}$, $\mu_-<0$, and hence the magnification of
the image which is inside the Einstein ring is negative. This means
that this image has its parity flipped with respect to the source.
The net magnification of flux in the two images is obtained by adding
the absolute magnifications,
\begin{equation}
  \mu = |\mu_+|+|\mu_-| = \frac{u^2+2}{u\sqrt{u^2+4}}\;.
\label{eq:2.29}
\end{equation}
When the source lies on the Einstein radius, we have
$\beta=\theta_{\rm E}$, $u=1$, and the total magnification becomes
\begin{equation}
  \mu = 1.17 + 0.17 = 1.34\;.
\label{eq:2.30}
\end{equation}

How can lensing by a point mass be detected? Unless the lens is
massive ($M>10^6\,M_\odot$ for a cosmologically distant source), the
angular separation of the two images is too small to be
resolved. However, even when it is not possible to see the multiple
images, the magnification can still be detected if the lens and source
move relative to each other, giving rise to lensing-induced time
variability of the source (Chang \& Refsdal 1979; Gott 1981). When
this kind of variability is induced by stellar mass lenses it is
referred to as {\em microlensing\/}. Microlensing was first observed
in the multiply-imaged QSO 2237$+$0305 (Irwin et al.\ 1989), and may
also have been seen in QSO 0957$+$561 (Schild \& Smith 1991; see also
Sect.\ \ref{NB:subsubsect3.7.4}). Paczy\'nski (1986b) had the
brilliant idea of using microlensing to search for so-called {\em
Massive Astrophysical Compact Halo Objects\/} (MACHOs, Griest 1991) in
the Galaxy. We discuss this topic in some depth in Sect.\
\ref{NB:subsect2.2}.

\subsection{Microlensing in the Galaxy}
\label{NB:subsect2.2}

\subsubsection{Basic Relations}
\label{NB:subsubsect2.2.1}

If the closest approach between a point mass lens and a source is
$\le\theta_{\rm E}$, the peak magnification in the lensing-induced
light curve is $\mu_{\rm max}\ge1.34$. A magnification of $1.34$
corresponds to a brightening by $0.32$ magnitudes, which is easily
detectable. Paczy\'nski (1986b) proposed monitoring millions of stars
in the LMC to look for such magnifications in a small fraction of the
sources. If enough events are detected, it should be possible to map
the distribution of stellar-mass objects in our Galaxy.

Perhaps the biggest problem with Paczy\'nski's proposal is that
monitoring a million stars or more primarily leads to the detection of
a huge number of variable stars. The intrinsically variable sources
must somehow be distinguished from stars whose variability is caused
by microlensing. Fortunately, the light curves of lensed stars have
certain tell-tale signatures --- the light curves are expected to be
symmetric in time and the magnification is expected to be achromatic
because light deflection does not depend on wavelength (but see the
more detailed discussion in Sect.\ \ref{NB:subsubsect2.2.4} below). In
contrast, intrinsically variable stars typically have asymmetric light
curves and do change their colors.

The expected time scale for microlensing-induced variations is given
in terms of the typical angular scale $\theta_{\rm E}$, the relative
velocity $v$ between source and lens, and the distance to the lens:
\begin{equation}
  t_0 = \frac{D_{\rm d}\theta_{\rm E}}{v} =
  0.214\,{\rm yr}\,\left(\frac{M}{M_\odot}\right)^{1/2}
  \left(\frac{D_{\rm d}}{10\,{\rm kpc}}\right)^{1/2}
  \left(\frac{D_{\rm ds}}{D_{\rm s}}\right)^{1/2}
  \left(\frac{200\,{\rm km}\,{\rm s}^{-1}}{v}\right)\;.
\label{eq:2.31}
\end{equation}
The ratio $D_{\rm ds}D_{\rm s}^{-1}$ is close to unity if the lenses
are located in the Galactic halo and the sources are in the LMC. If
light curves are sampled with time intervals between about an hour and
a year, MACHOs in the mass range $10^{-6}\,M_\odot$ to $10^2\,M_\odot$
are potentially detectable. Note that the measurement of $t_0$ in a
given microlensing event does not directly give $M$, but only a
combination of $M$, $D_{\rm d}$, $D_{\rm s}$, and $v$. Various ideas
to break this degeneracy have been discussed. Figure \ref{NB:fig2.8}
shows microlensing-induced light curves for six different minimum
separations $\Delta y=u_{\rm min}$ between the source and the lens.
The widths of the peaks are $\sim t_0$, and there is a direct
one-to-one mapping between $\Delta y$ and the maximum magnification at
the peak of the light curve. A microlensing light curve therefore
gives two observables, $t_0$ and $\Delta y$.

\begin{figure}[ht] %PP
%JO \begin{figure}
\begin{center}
\begin{minipage}[t]{0.6\hsize}
\end{minipage}
\end{center}
\caption{Microlensing-induced light curves for six minimum separations
  between the source and the lens, $\Delta y=0.1,~0.3,\ldots,1.1$. The
  separation is expressed in units of the Einstein radius.}
\label{NB:fig2.8}
\end{figure}

The chance of seeing a microlensing event is usually expressed in
terms of the optical depth, which is the probability that at any
instant of time a given star is within an angle $\theta_{\rm E}$ of a
lens. The optical depth is the integral over the number density
$n(D_{\rm d})$ of lenses times the area enclosed by the Einstein ring
of each lens, i.e.
\begin{equation}
  \tau = \frac{1}{\delta\omega}\,
  \int\,dV\,n(D_{\rm d})\,\pi\theta_{\rm E}^2\;,
\label{eq:2.32}
\end{equation}
where $dV=\delta\omega\,D_{\rm d}^2\,dD_{\rm d}$ is the volume of an
infinitesimal spherical shell with radius $D_{\rm d}$ which covers a
solid angle $\delta\omega$. The integral gives the solid angle covered
by the Einstein circles of the lenses, and the probability is obtained
upon dividing this quantity by the solid angle $\delta\omega$ which is
observed. Inserting equation (\ref{eq:2.22}) for the Einstein angle,
we obtain
\begin{equation}
  \tau = \int_0^{D_{\rm s}}\,\frac{4\pi G\rho}{c^2}\,
  \frac{D_{\rm d}D_{\rm ds}}{D_{\rm s}}\,dD_{\rm d} =
  \frac{4\pi G}{c^2}\,D_{\rm s}^2\,\int_0^1\,
  \rho(x)\,x(1-x)\,dx\;,
\label{eq:2.33}
\end{equation}
where $x\equiv D_{\rm d}D_{\rm s}^{-1}$ and $\rho$ is the mass density
of MACHOs. In writing (\ref{eq:2.33}), we have made use of the fact
that space is locally Euclidean, hence $D_{\rm ds}=D_{\rm s}-D_{\rm
d}$. If $\rho$ is constant along the line-of-sight, the optical depth
simplifies to
\begin{equation}
  \tau = \frac{2\pi}{3}\,\frac{G\rho}{c^2}\,D_{\rm s}^2\;.
\label{eq:2.34}
\end{equation}
It is important to note that the optical depth $\tau$ depends on the
{\em mass density\/} of lenses $\rho$ and not on their {\em mass\/}
$M$. The timescale of variability induced by microlensing, however,
does depend on the square root of the mass, as shown by eq.\
(\ref{eq:2.31}).

\subsubsection{Ongoing Galactic Microlensing Searches}
\label{NB:subsubsect2.2.2}

Paczy\'nski's suggestion that microlensing by compact objects in the
Galactic halo may be detected by monitoring the light curves of stars
in the LMC inspired several groups to set up elaborate searches for
microlensing events. Four groups, MACHO (Alcock et al.\ 1993), EROS
(Aubourg et al.\ 1993), OGLE (Udalski et al.\ 1992), and DUO (Alard
1995), are currently searching for microlensing-induced stellar
variability in the LMC (EROS, MACHO) as well as in the Galactic bulge
(DUO, MACHO, OGLE).

So far, about 100 microlensing events have been observed, and their
number is increasing rapidly. Most events have been seen toward the
Galactic bulge. The majority of events have been caused by single
lenses, and have light curves similar to those shown in Fig.\
\ref{NB:fig2.8}, but at least two events so far are due to binary
lenses. Strong lensing by binaries (defined as events where the source
crosses one or more caustics, see Fig.\ \ref{NB:fig2.9}) was estimated
by Mao \& Paczy\'nski (1991) to contribute about 10 per cent of all
events. Binary lensing is most easily distinguished from single-lens
events by characteristic double-peaked or asymmetric light curves;
Fig.\ \ref{NB:fig2.9} shows some typical examples.

\begin{figure}[ht] %PP
%JO \begin{figure}
\begin{center}
\begin{minipage}[t]{0.45\hsize}
\end{minipage}
\begin{minipage}[t]{0.45\hsize}
\end{minipage}
\end{center}
\caption{Left panel: A binary lens composed of two equal point
  masses. The critical curve is shown by the heavy line, and the
  corresponding caustic is indicated by the thin line with six
  cusps. (See Sect.\ \protect\ref{NB:subsubsect3.3.2} for a definition
  of critical curves and caustics.) Five source trajectories across
  this lens system are indicated. Right panel: Light curves
  corresponding to an extended source moving along the trajectories
  indicated in the left panel. Double-peaked features occur when the
  source comes close to both lenses.}
\label{NB:fig2.9}
\end{figure}

\begin{figure}[ht] %PP
%JO \begin{figure}
\begin{center}
\begin{minipage}[t]{0.9\hsize}
\end{minipage}
\end{center}
\caption{Light curve of the first binary microlensing event, OGLE~\#7
  (taken from the OGLE {\tt WWW} home page at {\tt
  http://www.astrouw.edu.pl/$\mathtt{\sim}$ftp/ogle/ogle.html}).}
\label{NB:fig2.10}
\end{figure}

The light curve of the first observed binary microlensing event,
OGLE~\#7, is shown in Fig.\ \ref{NB:fig2.10}.

\subsubsection{Early Results on Optical Depths}
\label{NB:subsubsect2.2.3}

Both the OGLE and MACHO collaborations have determined the
microlensing optical depth toward the Galactic bulge. The results are
\begin{equation}
  \tau = \left\{
  \begin{array}{ll}
    (3.3\pm1.2)\times10^{-6} & \hbox{(Paczy\'nski et al.\ 1994)} \\
    (3.9^{+1.8}_{-1.2})\times10^{-6} & \hbox{(Alcock et al.\ 1997)} \\
  \end{array}\right.\;.
\label{eq:2.35}
\end{equation}
Original theoretical estimates (Paczy\'nski 1991; Griest, Alcock, \&
Axelrod 1991) had predicted an optical depth below $10^{-6}$. Even
though this value was increased slightly by Kiraga \& Paczy\'nski
(1994) who realized the importance of lensing of background bulge
stars by foreground bulge stars (referred to as self-lensing of the
bulge), the measured optical depth is nevertheless very much higher
than expected. Paczy\'nski et al.\ (1994) suggested that a Galactic
bar which is approximately aligned with the line-of-sight toward the
Galactic bulge might explain the excess optical depth. Self-consistent
calculations of the bar by Zhao, Spergel, \& Rich (1995) and Zhao,
Rich, \& Spergel (1996) give $\tau\sim2\times10^{-6}$, which is within
one standard deviation of the observed value. However, using
COBE/DIRBE near-infrared data of the inner Galaxy and calibrating the
mass-to-light ratio with the terminal velocities of HI and CO clouds,
Bissantz et al.\ (1997) find a significantly lower optical depth,
$0.8\times10^{-6}\la\tau\la0.9\times10^{-6}$. Zhao \& Mao (1996)
describe how the shape of the Galactic bar can be inferred from
measuring the spatial dependence of the optical depth. Zhao et al.\
(1995) claim that the duration distribution of the bulge events
detected by OGLE is compatible with a roughly normal stellar mass
distribution.

In principle, moments of the mass distribution of microlensing objects
can be inferred from moments of the duration distribution of
microlensing events (De R\'ujula, Jetzer, \& Mass\'o 1991). Mao \&
Paczy\'nski (1996) have shown that a robust determination of mass
function parameters requires $\sim100$ microlensing events even if the
geometry of the microlens distribution and the kinematics are known.

Based on three events from their first year of data, of which two are
of only modest significance, Alcock et al.\ (1996) estimated the
optical depth toward the LMC to be
\begin{equation}
  \tau = 9^{+7}_{-5}\times10^{-8}\;,
\label{eq:2.36}
\end{equation}
in the mass range $10^{-4}\,M_\odot<M<10^{- 1}\,M_\odot$. This is too
small for the entire halo to be made of MACHOs in this mass range. At
the $95\%$ confidence level, the first-year data of the MACHO
collaboration rule out a contribution from MACHOs to the halo mass
$\ga40\%$ in the mass range $10^{-3}\,M_\odot\le
M\le10^{-2}\,M_\odot$, and $\ga60\%$ within $10^{-4}\,M_\odot\le
M\le10^{-1}\,M_\odot$. Sahu (1994) argued that all events can be due
to objects in the Galactic disk or the LMC itself. The EROS
collaboration, having better time resolution, is able to probe smaller
masses, $10^{-7}\,M_\odot\le M\le10^{-1}\,M_\odot$ (Aubourg et al.\
1995). The $95\%$ confidence level from the EROS data excludes a halo
fraction $\ga(20-30)\%$ in the mass range $10^{-7}\,M_\odot\le
M\le10^{-2}\,M_\odot$ (Ansari et al.\ 1996; Renault et al.\ 1997; see
also Roulet \& Mollerach 1997).

More recently, the MACHO group reported results from 2.3 years of
data. Based on 8 events, they now estimate the optical depth toward
the LMC to be
\begin{equation}
  \tau=2.9^{+1.4}_{-0.9}\times10^{-7}\;,
\label{eq:2.36a}
\end{equation}
and the halo fraction to be $0.45-1$ in the mass range
$0.2\,M_\odot\le M\le0.5\,M_\odot$ at 68\% confidence. Further, they
cannot reject, at the 99\% confidence level, the hypothesis that the
entire halo is made of MACHOs with masses $0.2\,M_\odot\le
M\le1\,M_\odot$ (Sutherland 1996). More data are needed before any
definitive conclusion can be reached on the contribution of MACHOs to
the halo.

\subsubsection{Other Interesting Discoveries}
\label{NB:subsubsect2.2.4}

In the simplest scenario of microlensing in the Galaxy, a single
point-like source is lensed by a single point mass which moves with
constant velocity relative to the source. The light curve observed
from such an event is time-symmetric and achromatic. At the low
optical depths that we expect in the Galaxy, and ignoring binaries,
microlensing events should not repeat since the probability that the
same star is lensed more than once is negligibly small.

In practice, the situation is more complicated and detailed
interpretations of observed light curves must account for some of the
complications listed below. The effects of binary lenses have already
been mentioned above. In the so-called {\em resonant\/} case, the
separation of the two lenses is comparable to their Einstein radii.
The light curve of such a lens system can have dramatic features such
as the double peaks shown in Figs.\ \ref{NB:fig2.9} and
\ref{NB:fig2.10}. At least two such events have been observed so far,
OGLE~\#7 (Udalski et al.\ 1994; Bennett et al.\ 1995) and DUO~\#2
(Alard, Mao, \& Guibert 1995). In the non-resonant case, the lenses
are well separated and act as almost independent lenses. Di Stefano \&
Mao (1996) estimated that a few per cent of all microlensing events
should ``repeat'' due to consecutive magnification of a star by the
two stars in a wide binary lens.

The sensitivity of microlensing searches to binaries may make this a
particularly powerful method to search for planets around distant
stars, as emphasized by Mao \& Paczy\'nski (1991) and Gould \& Loeb
(1992).

Multiple sources can give rise to various other complications. Since
the optical depth is low, microlensing searches are performed in
crowded fields where the number density of sources is high. Multiple
source stars which are closer to each other than $\sim1''$ appear as
single because they are not resolved. The Einstein radius of a solar
mass lens, on the other hand, is $\sim0\farcs001$ (cf.\ eq.\
\ref{eq:2.23}). Therefore, if the projected separation of two sources
is $<1''$ but $>0\farcs001$, a single lens affects only one of them at
a time. Several effects can then occur. First, the microlensing event
can apparently recur when the two sources are lensed individually
(Griest \& Hu 1992). Second, if the sources have different colors, the
event is chromatic because the color of the lensed star dominates
during the event (Udalski et al.\ 1994; Kamionkowski 1995; Buchalter,
Kamionkowski, \& Rich 1996). Third, the observed flux is a blend of
the magnified flux from the lensed component and the constant flux of
the unlensed components, and this leads to various biases (Di Stefano
\& Esin 1995; see also Alard \& Guibert 1997). A systematic method of
detecting microlensing in blended data has been proposed and is
referred to as ``pixel lensing'' (Crotts 1992; Baillon et al.\ 1993;
Colley 1995; Gould 1996).

Stars are not truly point-like. If a source is larger than the impact
parameter of a single lens or the caustic structure of a binary lens,
the finite source size modifies the light curve significantly (Gould
1994a; Nemiroff \& Wickramasinghe 1994; Witt \& Mao 1994; Witt 1995).

Finally, if the relative transverse velocity of the source, the lens,
and the observer is not constant during the event, the light curve
becomes time-asymmetric. The parallax effect due to the acceleration
of the Earth was predicted by Gould (1992b) and detected by Alcock et
al.\ (1995b). The detection of parallax provides an additional
observable which helps partially to break the degeneracy among $M$,
$v$, $D_{\rm d}$ and $D_{\rm ds}$ mentioned in Sect.\
\ref{NB:subsubsect2.2.1}.

Another method of breaking the degeneracy is via observations from
space. The idea of space measurements was suggested by Refsdal as
early as 1966 as a means to determine distances and masses of lenses
in the context of quasar lensing (Refsdal 1966b). Some obvious
benefits of space-based telescopes include absence of seeing and
access to wavebands like the UV or IR which are absorbed by the
Earth's atmosphere. The particular advantage of space observations for
microlensing in the Galaxy arises from the fact that the Einstein
radius of a sub-solar mass microlens in the Galactic halo is of order
$10^8\;{\rm km}$ and thus comparable to the AU, cf.\ eq.\
(\ref{eq:2.23}). Telescopes separated by $\sim1\;{\rm AU}$ would
therefore observe different light curves for the same microlensing
event. This additional information on the event can break the
degeneracy between the parameters defining the time scale $t_0$ (Gould
1994b). In the special (and rare) case of very high magnification when
the source is resolved during the event (Gould 1994a), all four
parameters may be determined.

Interesting discoveries can be expected from the various microlensing
``alert systems'' which have been recently set up (GMAN, Pratt 1996;
PLANET, Sackett 1996, Albrow et al.\ 1996). The goal of these
programs is to monitor ongoing microlensing events in almost real time
with very high time resolution. It should be possible to detect
anomalies in the microlensing lightcurves which are expected from the
complications listed in this section, or to obtain detailed
information (e.g.\ spectra, Sahu 1996) from objects while they are
being microlensed.

Jetzer (1994) showed that the microlensing optical depth toward the
Andromeda galaxy M~31 is similar to that toward the LMC,
$\tau\simeq10^{-6}$. Experiments to detect microlensing toward M~31
have recently been set up (e.g.\ Gondolo et al.\ 1997; Crotts \&
Tomaney 1996), and results are awaited.

\subsection{Extragalactic Microlenses}
\label{NB:subsubsect2.3}

\subsubsection{Point Masses in the Universe}
\label{NB:subsubsect2.3.1}

It has been proposed at various times that a significant fraction of
the dark matter in the universe may be in the form of compact masses.
These masses will induce various lensing phenomena, some of which are
very easily observed. The lack of evidence for these phenomena can
therefore be used to place useful limits on the fraction of the mass
in the universe in such objects (Press \& Gunn 1973).

Consider an Einstein-de Sitter universe with a constant comoving
number density of point lenses of mass $M$ corresponding to a cosmic
density parameter $\Omega_{\rm M}$. The optical depth for lensing of
sources at redshift $z_{\rm s}$ can be shown to be
\begin{eqnarray}
  \tau(z_{\rm s}) &=& 3\,\Omega_{\rm M}\,\left[
    \frac{(z_{\rm s}+2+2\sqrt{1+z_{\rm s}})\ln(1+z_{\rm s})}
    {z_{\rm s}}-4
  \right]\nonumber\\
  &\simeq&
  \Omega_{\rm M}\,\frac{z_{\rm s}^2}{4}
    \quad\hbox{for}\;z_{\rm s}\ll1\nonumber\\
  &\simeq&
  0.3\,\Omega_{\rm M}
    \quad\hbox{for}\;z_{\rm s}=2\;.\nonumber\\
\label{eq:2.37}
\end{eqnarray}
We see that the probability for lensing is $\sim\Omega_{\rm M}$ for
high-redshift sources (Press \& Gunn 1973). Hence the number of
lensing events in a given source sample directly measures the
cosmological density in compact objects.

In calculating the probability of lensing it is important to allow for
various selection effects. Lenses magnify the observed flux, and
therefore sources which are intrinsically too faint to be observed may
be lifted over the detection threshold. At the same time, lensing
increases the solid angle within which sources are observed so that
their number density in the sky is reduced (Narayan 1991). If there is
a large reservoir of faint sources, the increase in source number due
to the apparent brightening outweighs their spatial dilution, and the
observed number of sources is increased due to lensing. This
magnification bias (Turner 1980; Turner, Ostriker, \& Gott 1984;
Narayan \& Wallington 1993) can substantially increase the probability
of lensing for bright optical quasars whose number-count function is
steep.

\subsubsection{Current Upper Limits on $\Omega_{\rm M}$ in Point
Masses}
\label{NB:subsubsect2.3.2}

Various techniques have been proposed and applied to obtain limits on
$\Omega_{\rm M}$ over a broad range of lens masses (see Carr 1994 for
a review). Lenses with masses in the range $10^{10}<M/M_\odot<10^{12}$
will split images of bright QSOs by $0\farcs3-3''$. Such angular
splittings are accessible to optical observations; therefore, it is
easy to constrain $\Omega_{\rm M}$ in this mass range. The image
splitting of lenses with $10^6<M/M_\odot<10^8$ is on the order of
milliarcseconds and falls within the resolution domain of VLBI
observations of radio quasars (Kassiola, Kovner, \& Blandford
1991). The best limits presently are due to Henstock et al.\ (1995). A
completely different approach utilizes the differential time delay
between multiple images. A cosmological $\gamma$-ray burst, which is
gravitationally lensed will be seen as multiple repetitions of a
single event (Blaes \& Webster 1992). By searching the $\gamma$-ray
burst database for (lack of) evidence of repetitions, $\Omega_{\rm M}$
can be constrained over a range of masses which extends below the VLBI
range mentioned above. The region within QSOs where the broad emission
lines are emitted is larger than the region emitting the continuum
radiation. Lenses with $M\sim1\,M_\odot$ can magnify the continuum
relative to the broad emission lines and thereby reduce the observed
emission line widths. Lenses of still smaller masses cause apparent
QSO variability, and hence from observations of the variability an
upper limit to $\Omega_{\rm M}$ can be derived. Finally, the time
delay due to lenses with very small masses can be such that the light
beams from multiply imaged $\gamma$-ray bursts interfere so that the
observed burst spectra should show interference patterns. Table
\ref{NB:tab2.1} summarizes these various techniques and gives the most
recent results on $\Omega_{\rm M}$.

\begin{table}
  \caption{Summary of techniques to constrain $\Omega_{\rm M}$ in
    point masses, along with the current best limits.}
\medskip %PP
  \begin{center}
  \begin{tabular}{|l|lrr|}
  \hline\hline
    Technique & References & Mass Range & Limit on \\
    && $M_\odot$ & $\Omega_{\rm M}$ \\
  \hline\hline
    Image doubling of & Surdej et al.\ (1993) &
    $10^{10}-10^{12}$ & $<0.02$ \\
    bright QSOs &&&\\
  \hline
    Doubling of VLBI & Kassiola et al.\ (1991) &
    $10^6-10^8$ & $<0.05$ \\
    compact sources & Henstock et al.\ (1995) &&\\
  \hline
    Echoes from $\gamma$-ray & Nemiroff et al.\ (1993) &
    $10^{6.5}-10^{8.1}$ & $\la1$ \\
    bursts & & & excluded \\
    & Nemiroff et al.\ (1994) & $(10^3\to)$ & null result \\
  \hline
    Diff. magnification & Canizares (1982) &
    $10^{-1}-20$ & $<0.1$ \\
    of QSO continuum vs. & Dalcanton et al.\ (1994) &
    $10^{-3}-60$ & $<0.2$ \\
    broad emission lines &&&\\
  \hline
    Quasar variability & Schneider (1993) &
    $10^{-3}-10^{-2}$ & $<0.1$ \\
  \hline
    Femtolensing of $\gamma$-ray & Gould (1992a) &
    $10^{-17}-10^{-13}$ & -- \\
    bursts & Stanek et al.\ (1993) & & -- \\
  \hline\hline
  \end{tabular}
  \end{center}
\label{NB:tab2.1}
\end{table}

As Table \ref{NB:tab2.1} shows, we can eliminate $\Omega_{\rm M}\sim1$
in virtually all astrophysically plausible mass ranges. The limits are
especially tight in the range $10^6<M/M_\odot<10^{12}$, where
$\Omega_{\rm M}$ is constrained to be less than a few per cent.

\subsubsection{Microlensing in QSO 2237$+$0305}
\label{NB:subsubsect2.3.3}

Although the Galactic microlensing projects described earlier have
developed into one of the most exciting branches of gravitational
lensing, the phenomenon of microlensing was in fact first detected in
a cosmological source, the quadruply-imaged QSO 2237$+$0305 (Irwin et
al.\ 1989; Corrigan et al.\ 1991; Webster et al.\ 1991; {\O}stensen et
al.\ 1996). The lensing galaxy in QSO 2237$+$0305 is a spiral at a
redshift of $0.04$ (Huchra et al.\ 1985). The four quasar images are
almost symmetrically located in a cross-shaped pattern around the
nucleus of the galaxy; hence the system has been named the ``Einstein
Cross''. Uncorrelated flux variations have been observed in QSO
2237$+$0305, possibly in all four images, and these variations provide
evidence for microlensing due to stars in the lensing galaxy. Figure
\ref{NB:fig2.11} shows the light curves of the four images.

\begin{figure}[ht] %PP
%JO \begin{figure}
\begin{center}
\begin{minipage}[t]{0.6\hsize}
\end{minipage}
\end{center}
\caption{Light curves of the four images in the ``Einstein Cross'' QSO
  2237$+$0305 since August 1990 (from {\O}stensen et al.\ 1996)}
\label{NB:fig2.11}
\end{figure}

The interpretation of the microlensing events in QSO 2237$+$0305 is
much less straightforward than in the case of microlensing in the
Galaxy. When a distant galaxy forms multiple images, the surface mass
density at the locations of the images is of order the critical
density, $\Sigma_{\rm cr}$. If most of the local mass is made of stars
or other massive compact objects (as is likely in the case of QSO
2237$+$0305 since the four images are superposed on the bulge of the
lensing spiral galaxy), the optical depth to microlensing approaches
unity. In such a case, the mean projected separation of the stars is
comparable to or smaller than their Einstein radii, and the effects of
the various microlenses cannot be considered
independently. Complicated caustic networks arise (Paczy\'nski 1986a;
Schneider \& Weiss 1987; Wambsganss 1990), and the observed light
curves have to be analyzed statistically on the basis of numerical
simulations. A new and elegant method to compute microlensing light
curves of point sources was introduced by Witt (1993).

Two important conclusions have been drawn from the microlensing events
in QSO 2237$+$0305. First, it has been shown that the continuum
emitting region in QSO 2237$+$0305 must have a size $\sim10^{15}$~cm
(Wambsganss, Paczy\'nski, \& Schneider 1990; Rauch \& Blandford 1991)
in order to produce the observed amplitude of magnification
fluctuations. This is the most direct and stringent limit yet on the
size of an optical QSO. Second, it appears that the mass spectrum of
microlenses in the lensing galaxy is compatible with a normal mass
distribution similar to that observed in our own Galaxy (Seitz,
Wambsganss, \& Schneider 1994).

\section{Lensing by Galaxies}
\label{NB:sect3}

Lensing by point masses, the topic we have considered so far, is
particularly straightforward because of the simplicity of the lens.
When we consider galaxy lenses we need to allow for the distributed
nature of the mass, which is usually done via a parameterized model.
The level of complexity of the model is dictated by the application at
hand.

\subsection{Lensing by a Singular Isothermal Sphere}
\label{NB:subsect3.1}

A simple model for the mass distribution in galaxies assumes that the
stars and other mass components behave like particles of an ideal gas,
confined by their combined, spherically symmetric gravitational
potential. The equation of state of the ``particles'', henceforth
called stars for simplicity, takes the form
\begin{equation}
  p = \frac{\rho\,kT}{m}\;,
\label{eq:3.1}
\end{equation}
where $\rho$ and $m$ are the mass density and the mass of the stars.
In thermal equilibrium, the temperature $T$ is related to the
one-dimensional velocity dispersion $\sigma_v$ of the stars through
\begin{equation}
  m\sigma_v^2 = kT\;.
\label{eq:3.2}
\end{equation}
The temperature, or equivalently the velocity dispersion, could in
general depend on radius $r$, but it is usually assumed that the
stellar gas is isothermal, so that $\sigma_v$ is constant across the
galaxy. The equation of hydrostatic equilibrium then gives
\begin{equation}
  \frac{p'}{\rho} = -\frac{GM(r)}{r^2}\;,\quad
  M'(r) = 4\pi\,r^2\,\rho\;,
\label{eq:3.3}
\end{equation}
where $M(r)$ is the mass interior to radius $r$, and primes denote
derivatives with respect to $r$. A particularly simple solution of
eqs.\ (\ref{eq:3.1}) through (\ref{eq:3.3}) is
\begin{equation}
  \rho(r) = \frac{\sigma_v^2}{2\pi G}\,\frac{1}{r^2}\;.
\label{eq:3.4}
\end{equation}
This mass distribution is called the {\em singular isothermal
sphere\/}. Since $\rho\propto r^{-2}$, the mass $M(r)$ increases
$\propto r$, and therefore the rotational velocity of test particles
in circular orbits in the gravitational potential is
\begin{equation}
  v_{\rm rot}^2(r) = \frac{G\,M(r)}{r} =
  2\,\sigma_v^2 = \hbox{constant}\;.
\label{eq:3.5}
\end{equation}
The flat rotation curves of galaxies are naturally reproduced by this
model.

Upon projecting along the line-of-sight, we obtain the surface mass
density
\begin{equation}
  \Sigma(\xi) = \frac{\sigma_v^2}{2G}\,\frac{1}{\xi}\;,
\label{eq:3.6}
\end{equation}
where $\xi$ is the distance from the center of the two-dimensional
profile. Referring to eq.\ (\ref{eq:2.12}), we immediately obtain the
deflection angle
\begin{equation}
  \hat\alpha = 4\pi\,\frac{\sigma_v^2}{c^2} = (1\farcs4)\,
  \left(\frac{\sigma_v}{220\,{\rm km}\,{\rm s}^{-1}}\right)^2\;,
\label{eq:3.7}
\end{equation}
which is independent of $\xi$ and points toward the center of the
lens. The Einstein radius of the singular isothermal sphere follows
from eq.\ (\ref{eq:2.21}),
\begin{equation}
  \theta_{\rm E} = 4\pi\,\frac{\sigma_v^2}{c^2}\,
  \frac{D_{\rm ds}}{D_{\rm s}} =
  \hat\alpha\,\frac{D_{\rm ds}}{D_{\rm s}} = \alpha\;.
\label{eq:3.8}
\end{equation}
Due to circular symmetry, the lens equation is essentially
one-dimensional. Multiple images are obtained only if the source lies
inside the Einstein ring, i.e. if $\beta<\theta_{\rm E}$. When this
condition is satisfied, the lens equation has the two solutions
\begin{equation}
  \theta_\pm = \beta\pm\theta_{\rm E}\;.
\label{eq:3.9}
\end{equation}
The images at $\theta_\pm$, the source, and the lens all lie on a
straight line. Technically, a third image with zero flux is located at
$\theta=0$. This third image acquires a finite flux if the singularity
at the center of the lens is replaced by a core region with a finite
density.

The magnifications of the two images follow from eq.\ (\ref{eq:2.27}),
\begin{equation}
  \mu_\pm = \frac{\theta_\pm}{\beta} =
  1\pm\frac{\theta_{\rm E}}{\beta} =
  \left(1\mp\frac{\theta_{\rm E}}{\theta_\pm}\right)^{-1}\;.
\label{eq:3.10}
\end{equation}
If the source lies outside the Einstein ring, i.e.\ if
$\beta>\theta_{\rm E}$, there is only one image at
$\theta=\theta_+=\beta+\theta_{\rm E}$.

\subsection{Effective Lensing Potential}
\label{NB:subsect3.2}

Before proceeding to more complicated galaxy lens models, it is useful
to develop the formalism a little further. Let us define a scalar
potential $\psi(\vec\theta)$ which is the appropriately scaled,
projected Newtonian potential of the lens,
\begin{equation}
  \psi(\vec\theta) = \frac{D_{\rm ds}}{D_{\rm d}D_{\rm s}}\,
  \frac{2}{c^2}\,\int\,\Phi(D_{\rm d}\vec\theta,z)\,dz\;.
\label{eq:3.11}
\end{equation}
The derivatives of $\psi$ with respect to $\vec\theta$ have convenient
properties. The gradient of $\psi$ with respect to $\theta$ is the
deflection angle,
\begin{equation}
  \vec\nabla_\theta\psi = D_{\rm d}\vec\nabla_\xi\psi =
  \frac{2}{c^2}\,\frac{D_{\rm ds}}{D_{\rm s}}\,
  \int\,\vec\nabla_\perp\Phi\,dz = \vec\alpha\;,
\label{eq:3.12}
\end{equation}
while the Laplacian is proportional to the surface-mass density
$\Sigma$,
\begin{equation}
  \nabla^2_\theta\psi =
  \frac{2}{c^2}\,\frac{D_{\rm d}D_{\rm ds}}{D_{\rm s}}\,
  \int\,\nabla_\xi^2\Phi\,dz =
  \frac{2}{c^2}\,\frac{D_{\rm d}D_{\rm ds}}{D_{\rm s}}\cdot
  4\pi G\,\Sigma =
  2\,\frac{\Sigma(\vec\theta)}{\Sigma_{\rm cr}} \equiv
  2\kappa(\vec\theta)\;,
\label{eq:3.13}
\end{equation}
where Poisson's equation has been used to relate the Laplacian of
$\Phi$ to the mass density. The surface mass density scaled with its
critical value $\Sigma_{\rm cr}$ is called the {\em convergence\/}
$\kappa(\vec\theta)$. Since $\psi$ satisfies the two-dimensional
Poisson equation $\nabla^2_\theta\psi=2\kappa$, the effective lensing
potential can be written in terms of $\kappa$
\begin{equation}
  \psi(\vec\theta) = \frac{1}{\pi}\,
  \int\,\kappa(\vec\theta')\ln|\vec\theta-\vec\theta'|\,
  d^2\theta'\;.
\label{eq:3.14}
\end{equation}
As mentioned earlier, the deflection angle is the gradient of $\psi$,
hence
\begin{equation}
  \vec\alpha(\vec\theta) = \vec\nabla\psi =
  \frac{1}{\pi}\,\int\,\kappa(\vec\theta')\,
  \frac{\vec\theta-\vec\theta'}{|\vec\theta-\vec\theta'|^2}\,
  d^2\theta'\;,
\label{eq:3.15}
\end{equation}
which is equivalent to eq.\ (\ref{eq:2.11}) if we account for the
definition of $\Sigma_{\rm cr}$ given in eq.\ (\ref{eq:2.18}).

The local properties of the lens mapping are described by its Jacobian
matrix ${\cal A}$,
\begin{equation}
  {\cal A} \equiv \frac{\partial\vec\beta}{\partial\vec\theta} =
  \left(\delta_{ij} -
  \frac{\partial\alpha_i(\vec\theta)}{\partial\theta_j}\right) =
  \left(\delta_{ij} - \frac{\partial^2\psi(\vec\theta)}
  {\partial\theta_i\partial\theta_j}\right) = {\cal M}^{-1} \;.
\label{eq:3.16}
\end{equation}
As we have indicated, ${\cal A}$ is nothing but the inverse of the
magnification tensor ${\cal M}$. The matrix ${\cal A}$ is therefore
also called the inverse magnification tensor. The local solid-angle
distortion due to the lens is given by the determinant of ${\cal
A}$. A solid-angle element $\delta\beta^2$ of the source is mapped to
the solid-angle element of the image $\delta\theta^2$, and so the
magnification is given by
\begin{equation}
  \frac{\delta\theta^2}{\delta\beta^2} = \det{\cal M} =
  \frac{1}{\det{\cal A}}\;.
\label{eq:3.17}
\end{equation}
This expression is the appropriate generalization of eq.\
(\ref{eq:2.27}) when there is no symmetry.

Equation (\ref{eq:3.16}) shows that the matrix of second partial
derivatives of the potential $\psi$ (the Hessian matrix of $\psi$)
describes the deviation of the lens mapping from the identity
mapping. For convenience, we introduce the abbreviation
\begin{equation}
  \frac{\partial^2\psi}{\partial\theta_i\partial\theta_j} \equiv
  \psi_{ij}\;.
\label{eq:3.18}
\end{equation}
Since the Laplacian of $\psi$ is twice the convergence, we have
\begin{equation}
  \kappa = \frac{1}{2}\,(\psi_{11}+\psi_{22}) =
  \frac{1}{2}\,\tr\psi_{ij}\;.
\label{eq:3.19}
\end{equation}
Two additional linear combinations of $\psi_{ij}$ are important, and
these are the components of the {\em shear\/} tensor,
\begin{eqnarray}
  \gamma_1(\vec\theta) &=&
  \frac{1}{2}\left(\psi_{11}-\psi_{22}\right)
  \equiv \gamma(\vec\theta)\cos\left[2\phi(\vec\theta)\right]\;,
  \nonumber\\
  \gamma_2(\vec\theta) &=& \psi_{12} = \psi_{21}
  \equiv \gamma(\vec\theta)\sin\left[2\phi(\vec\theta)\right]\;.
  \nonumber\\
\label{eq:3.20}
\end{eqnarray}
With these definitions, the Jacobian matrix can be written
\begin{eqnarray}
  {\cal A} &=& \pmatrix{1-\kappa-\gamma_1 & -\gamma_2 \cr
                        -\gamma_2 & 1-\kappa+\gamma_1 \cr}
  \nonumber\\
  &=& (1-\kappa)\,
  \pmatrix{1 & 0 \cr 0 & 1 \cr} - \gamma\,
  \pmatrix{\cos2\phi & \sin2\phi \cr \sin2\phi & -\cos2\phi\cr}\;.
  \nonumber\\
\label{eq:3.21}
\end{eqnarray}
The meaning of the terms convergence and shear now becomes intuitively
clear. Convergence acting alone causes an isotropic focusing of light
rays, leading to an isotropic magnification of a source. The source is
mapped onto an image with the same shape but larger size. Shear
introduces anisotropy (or astigmatism) into the lens mapping; the
quantity $\gamma=(\gamma_1^2+\gamma_2^2)^{1/2}$ describes the
magnitude of the shear and $\phi$ describes its orientation. As shown
in Fig.\ \ref{NB:fig3.1}, a circular source of unit radius becomes, in
the presence of both $\kappa$ and $\gamma$, an elliptical image with
major and minor axes
\begin{equation}
  (1-\kappa-\gamma)^{-1}\;,\quad(1-\kappa+\gamma)^{-1}\;.
\label{eq:3.22}
\end{equation}
The magnification is
\begin{equation}
  \mu = \det{\cal M} =
  \frac{1}{\det{\cal A}} = \frac{1}{[(1-\kappa)^2-\gamma^2]}\;.
\label{eq:3.23}
\end{equation}
Note that the Jacobian ${\cal A}$ is in general a function of position
$\vec\theta$.

\begin{figure}[ht] %PP
%JO \begin{figure}
\begin{center}
\begin{minipage}[t]{0.6\hsize}
\end{minipage}
\end{center}
\caption{Illustration of the effects of convergence and shear on a
  circular source. Convergence magnifies the image isotropically, and
  shear deforms it to an ellipse.}
\label{NB:fig3.1}
\end{figure}

\subsection{Gravitational Lensing via Fermat's Principle}
\label{NB:subsect3.3}

\subsubsection{The Time-Delay Function}
\label{NB:subsubsect3.3.1}

The lensing properties of model gravitational lenses are especially
easy to visualize by application of Fermat's principle of geometrical
optics (Nityananda 1984, unpublished; Schneider 1985; Blandford \&
Narayan 1986; Nityananda \& Samuel 1992). From the lens equation
(\ref{eq:2.15}) and the fact that the deflection angle is the gradient
of the effective lensing potential $\psi$, we obtain
\begin{equation}
  (\vec\theta-\vec\beta)-\vec\nabla_\theta\psi = 0\;.
\label{eq:3.24}
\end{equation}
This equation can be written as a gradient,
\begin{equation}
  \vec\nabla_\theta
  \left[\frac{1}{2}(\vec\theta-\vec\beta)^2-\psi\right]=0\;.
\label{eq:3.25}
\end{equation}
The physical meaning of the term in square brackets becomes more
obvious by considering the time-delay function,
\begin{eqnarray}
  t(\vec\theta) &=& \frac{(1+z_{\rm d})}{c}\,
  \frac{D_{\rm d}D_{\rm s}}{D_{\rm ds}}\,
  \left[\frac{1}{2}(\vec\theta-\vec\beta)^2-
  \psi(\vec\theta)\right]
  \nonumber\\
  &=& t_{\rm geom} + t_{\rm grav}\;.\nonumber\\
\label{eq:3.26}
\end{eqnarray}
The term $t_{\rm geom}$ is proportional to the square of the angular
offset between $\vec\beta$ and $\vec\theta$ and is the time delay due
to the extra path length of the deflected light ray relative to an
unperturbed null geodesic. The coefficient in front of the square
brackets ensures that the quantity corresponds to the time delay as
measured by the observer. The second term $t_{\rm grav}$ is the time
delay due to gravity and is identical to the Shapiro delay introduced
in eq.\ (\ref{eq:2.4}), with an extra factor of $(1+z_{\rm d})$ to
allow for time stretching. Equations (\ref{eq:3.25}) and
(\ref{eq:3.26}) imply that images satisfy the condition
$\vec\nabla_\theta t(\vec\theta)=0$ (Fermat's Principle).

\begin{figure}[ht] %PP
%JO \begin{figure}
\begin{center}
\begin{minipage}[t]{0.6\hsize}
\end{minipage}
\end{center}
\caption{Geometric, gravitational, and total time delay of a
  circularly symmetric lens for a source that is slightly offset from
  the symmetry axis. The dotted line shows the location of the center
  of the lens, and $\beta$ shows the position of the source. Images
  are located at points where the total time delay function is
  stationary. The image positions are marked with dots in the bottom
  panel.}
\label{NB:fig3.2}
\end{figure}

In the case of a circularly symmetric deflector, the source, the lens
and the images have to lie on a straight line on the sky. Therefore,
it is sufficient to consider the section along this line of the time
delay function. Figure \ref{NB:fig3.2} illustrates the geometrical and
gravitational time delays for this case. The top panel shows $t_{\rm
geom}$ for a slightly offset source. The curve is a parabola centered
on the position of the source. The central panel displays $t_{\rm
grav}$ for an isothermal sphere with a softened core. This curve is
centered on the lens. The bottom panel shows the total
time-delay. According to the above discussion images are located at
stationary points of $t(\theta)$. For the case shown in Fig.\
\ref{NB:fig3.2} there are three stationary points, marked by dots, and
the corresponding values of $\theta$ give the image positions.

\subsubsection{Properties of the Time-Delay Function}
\label{NB:subsubsect3.3.2}

In the general case it is necessary to consider image locations in the
two-dimensional space of $\vec\theta$, not just on a line. The images
are then located at those points $\vec\theta_i$ where the
two-dimensional time-delay surface $t(\vec\theta)$ is stationary.
This is {\em Fermat's Principle\/} in geometrical optics, which states
that the actual trajectory followed by a light ray is such that the
light-travel time is stationary relative to neighboring trajectories.
The time-delay surface $t(\vec\theta)$ has a number of useful
properties.

\begin{itemize}

\item[1] The height difference between two stationary points on
$t(\vec\theta)$ gives the relative time delay between the
corresponding images. Any variability in the source is observed first
in the image corresponding to the lowest point on the surface,
followed by the extrema located at successively larger values of
$t$. In Fig.\ \ref{NB:fig3.2} for instance, the first image to vary is
the one that is farthest from the center of the lens. Although Fig.\
\ref{NB:fig3.2} corresponds to a circularly symmetric lens,this
property usually carries over even for lenses that are not perfectly
circular. Thus, in QSO 0957$+$561, we expect the A image, which is
$\sim5''$ from the lensing galaxy, to vary sooner than the B image,
which is only $\sim1''$ from the center. This is indeed observed (for
recent optical and radio light curves of QSO 0957+561 see Schild \&
Thomson 1993; Haarsma et al.\ 1996, 1997; Kundi\'c et al.\ 1996).

\item[2] There are three types of stationary points of a
two-dimensional surface: minima, saddle points, and maxima. The nature
of the stationary points is characterized by the eigenvalues of the
Hessian matrix of the time-delay function at the location of the
stationary points,
\begin{equation}
  {\cal T} =
  \frac{\partial^2t(\vec\theta)}{\partial\theta_i\partial\theta_j}
  \propto \left(\delta_{ij}-\psi_{ij}\right) = {\cal A} \;.
\label{eq:3.27}
\end{equation}
The matrix ${\cal T}$ describes the local curvature of the time-delay
surface. If both eigenvalues of ${\cal T}$ are positive,
$t(\vec\theta)$ is curved ``upward'' in both coordinate directions,
and the stationary point is a minimum. If the eigenvalues of ${\cal
T}$ have opposite signs we have a saddle point, and if both
eigenvalues of ${\cal T}$ are negative, we have a maximum.
Correspondingly, we can distinguish three types of images. Images of
type I arise at minima of $t(\vec\theta)$ where $\det{\cal A}>0$ and
$\tr{\cal A}>0$. Images of type II are formed at saddle points of
$t(\vec\theta)$ where the eigenvalues have opposite signs, hence
$\det{\cal A}<0$. Images of type III are located at maxima of
$t(\vec\theta)$ where both eigenvalues are negative and so $\det{\cal
A}>0$ and $\tr{\cal A}<0$.

\item[3] Since the magnification is the inverse of $\det{\cal A}$,
images of types I and III have positive magnification and images of
type II have negative magnification. The interpretation of a negative
$\mu$ is that the parity of the image is reversed. A little thought
shows that the three images shown in Fig.\ \ref{NB:fig3.2} correspond,
from the left, to a saddle-point, a maximum and a minimum,
respectively. The images A and B in QSO 0957$+$561 correspond to the
images on the right and left in this example, and ought to represent a
minimum and a saddle-point respectively in the time delay
surface. VLBI observations do indeed show the expected reversal of
parity between the two images (Gorenstein et al.\ 1988).

\begin{figure}[ht] %PP
%JO \begin{figure}
\begin{center}
\begin{minipage}[t]{0.6\hsize}
\end{minipage}
\end{center}
\caption{The time delay function of a circularly symmetric lens for a
  source exactly behind the lens (top panel), a source offset from the
  lens by a moderate angle (center panel) and a source offset by a
  large angle (bottom panel).}
\label{NB:fig3.3}
\end{figure}

\item[4] The curvature of $t(\vec\theta)$ measures the inverse
magnification. When the curvature of $t(\vec\theta)$ along one
coordinate direction is small, the image is strongly magnified along
that direction, while if $t(\vec\theta)$ has a large curvature the
magnification is small. Figure \ref{NB:fig3.3} displays the time-delay
function of a typical circularly symmetric lens and a source on the
symmetry axis (top panel), a slightly offset source (central panel),
and a source with a large offset (bottom panel). If the separation
between source and lens is large, only one image is formed, while if
the source is close to the lens three images are formed. Note that, as
the source moves, two images approach each other, merge and vanish. It
is easy to see that, quite generally, the curvature of $t(\theta)$
goes to zero as the images approach each other; in fact, the curvature
varies as $\Delta\theta^{-1}$. Thus, we expect that the brightest
image configurations are obtained when a pair of images are close
together, just prior to merging. The lines in $\vec\theta$-space on
which images merge are referred to as {\em critical lines\/}, while
the corresponding source positions in $\vec\beta$-space are called
{\em caustics\/}. Critical lines and caustics are important because
(i) they highlight regions of high magnification, and (ii) they
demarcate regions of different image multiplicity. (The reader is
referred to Blandford \& Narayan 1986 and Erdl \& Schneider 1992 for
more details.)

\item[5] When the source is far from the lens, we expect only a single
image, corresponding to a minimum of the time delay surface. New
extrema are always created in pairs (e.g.\ Fig.\
\ref{NB:fig3.3}). Therefore, the total number of extrema, and thus the
number of images of a generic (non-singular) lens, is odd (Burke
1981).

\end{itemize}

\subsection{Circularly Symmetric Lens Models}
\label{NB:subsect3.4}

Table \ref{NB:tab3.1} compiles formulae for the effective lensing
potential and deflection angle of four commonly used circularly
symmetric lens models; point mass, singular isothermal sphere,
isothermal sphere with a softened core, and constant density sheet. In
addition, one can have more general models with non-isothermal radial
profiles, e.g. density varying as radius to a power other than $-2$.

\begin{table}
  \caption{Examples of circularly symmetric lenses. The effective
  lensing potential $\psi(\theta)$ and the deflection angle
  $\alpha(\theta)$ are given. The core radius of the softened
  isothermal sphere is $\theta_{\rm c}$.}
\medskip %PP
  \begin{center}
  \begin{tabular}{|l|cc|}
  \hline\hline
    Lens Model & $\psi(\theta)$ & $\alpha(\theta)$ \\
  \hline\hline
    Point mass &
    $\displaystyle{\frac{D_{\rm ds}}{D_{\rm s}}\,
      \frac{4GM}{D_{\rm d}c^2}\,\ln|\theta|}$ &
    $\displaystyle{\frac{D_{\rm ds}}{D_{\rm s}}\,
      \frac{4GM}{c^2D_{\rm d}|\theta|}}$ \\
  \hline
    Singular isothermal sphere &
    $\displaystyle{\frac{D_{\rm ds}}{D_{\rm s}}\,
      \frac{4\pi\sigma^2}{c^2}\,|\theta|}$ &
    $\displaystyle{\frac{D_{\rm ds}}{D_{\rm s}}\,
      \frac{4\pi\sigma^2}{c^2}}$ \\
  \hline
    Softened isothermal sphere &
    $\displaystyle{\frac{D_{\rm ds}}{D_{\rm s}}\,
      \frac{4\pi\sigma^2}{c^2}\,
      \left(\theta_{\rm c}^2+\theta^2\right)^{1/2}}$ &
    $\displaystyle{\frac{D_{\rm ds}}{D_{\rm s}}\,
      \frac{4\pi\sigma^2}{c^2}\,\frac{\theta}
      {\left(\theta_{\rm c}^2+\theta^2\right)^{1/2}}}$ \\
  \hline
    Constant density sheet &
    $\displaystyle{\frac{\kappa}{2}\,\theta^2}$ &
    $\displaystyle{\kappa|\theta|}$ \\
  \hline\hline
  \end{tabular}
  \end{center}
\label{NB:tab3.1}
\end{table}

The gravitational time-delay functions $t_{\rm
grav}(\theta)\propto-\psi(\theta)$ of the models in Table
\ref{NB:tab3.1} are illustrated in Fig.\ \ref{NB:fig3.4}. Note that
the four potentials listed in Tab.\ \ref{NB:tab3.1} all are divergent
for $\theta\to\infty$. (Although the three-dimensional potential of
the point mass drops $\propto r^{-1}$, its projection along the
line-of-sight diverges logarithmically.) The divergence is, however,
not serious since images always occur at finite $\theta$ where the
functions are well-behaved.

\begin{figure}[ht] %PP
%JO \begin{figure}
\begin{center}
\begin{minipage}[t]{0.6\hsize}
\end{minipage}
\end{center}
\caption{Gravitational time-delay functions for the four circularly
  symmetric effective potentials listed in Tab.\
  \protect\ref{NB:tab3.1}. (a) point mass; (b) singular isothermal
  sphere; (c) softened isothermal sphere with core radius $\theta_{\rm
  c}$; (d) constant density sheet.}
\label{NB:fig3.4}
\end{figure}

The image configurations produced by a circularly symmetric lens are
easily discovered by drawing time delay functions $t(\theta)$ as in
Fig.\ \ref{NB:fig3.3} corresponding to various offsets of the source
with respect to the lens. Figures \ref{NB:fig3.5} and \ref{NB:fig3.6}
show typical image configurations. The right halves of the figures
display the source plane, and the left halves show the image
configuration in the lens plane. Since ${\cal A}$ is a $2\times2$
matrix, a typical circularly symmetric lens has two critical lines
where $\det{\cal A}$ vanishes, and two corresponding caustics in the
source plane. The caustic of the inner critical curve is a circle
while the caustic of the outer critical curve degenerates to a
critical point because of the circular symmetry of the lens. A source
which is located outside the outermost caustic has a single
image. Upon each caustic crossing, the image number changes by two,
indicated by the numbers in Fig.\ \ref{NB:fig3.5}. The source shown as
a small rectangle in the right panel of Fig.\ \ref{NB:fig3.5} has
three images as indicated in the left panel. Of the three image, the
innermost one is usually very faint; in fact, this image vanishes if
the lens has a singular core (the curvature of the time delay function
then becomes infinite) as in the point mass or the singular isothermal
sphere.

\begin{figure}[ht] %PP
%JO \begin{figure}
\begin{center}
\begin{minipage}[t]{0.6\hsize}
\end{minipage}
\end{center}
\caption{Imaging of a point source by a non-singular,
  circularly-symmetric lens. Left: image positions and critical lines;
  right: source position and corresponding caustics.}
\label{NB:fig3.5}
\end{figure}

Figure \ref{NB:fig3.6} shows the images of two extended sources lensed
by the same model as in Fig.\ \ref{NB:fig3.5}. One source is located
close to the point-like caustic in the center of the lens. It is
imaged onto the two long, tangentially oriented arcs close to the
outer critical curve and the very faint image at the lens center. The
other source is located on the outer caustic and forms a radially
elongated image which is composed of two merging images, and a third
tangentially oriented image outside the outer caustic. Because of the
image properties, the outer critical curve is called {\em
tangential\/}, and the inner critical curve is called {\em radial\/}.

\begin{figure}[ht] %PP
%JO \begin{figure}
\begin{center}
\begin{minipage}[t]{0.6\hsize}
\end{minipage}
\end{center}
\caption{Imaging of an extended source by a non-singular
  circularly-symmetric lens. A source close to the point caustic at
  the lens center produces two tangentially oriented arc-like images
  close to the outer critical curve, and a faint image at the lens
  center. A source on the outer caustic produces a radially elongated
  image on the inner critical curve, and a tangentially oriented image
  outside the outer critical curve. Because of these image properties,
  the outer and inner critical curves are called {\em tangential\/}
  and {\em radial\/}, respectively.}
\label{NB:fig3.6}
\end{figure}

\subsection{Non-Circularly-Symmetric Lens Models}
\label{NB:subsect3.5}

A circularly symmetric lens model is much too idealized and is
unlikely to describe real galaxies. Therefore, considerable work has
gone into developing non-circularly symmetric models. The breaking of
the symmetry leads to qualitatively new image configurations (see
Grossman \& Narayan 1988; Narayan \& Grossman 1989; Blandford et al.\
1989).

\subsubsection{Elliptical Galaxy Model}
\label{NB:subsubsect3.5.1}

To describe an elliptical galaxy lens, we should ideally consider
elliptical isodensity contours. A straightforward generalization of
the isothermal sphere with finite core gives
\begin{equation}
  \Sigma(\theta_1,\theta_2) = \frac{\Sigma_0}
  {\left[\theta_{\rm c}^2 + (1-\epsilon)\theta_1^2 +
  (1+\epsilon)\theta_2^2\right]^{1/2}}\;,
\label{eq:3.28}
\end{equation}
where $\theta_1$, $\theta_2$ are orthogonal coordinates along the
major and minor axes of the lens measured from the center. The
potential $\psi(\theta_1,\theta_2)$ corresponding to this density
distribution has been calculated by Kassiola \& Kovner (1993) but is
somewhat complicated. For the specific case when the core radius
$\theta_{\rm c}$ vanishes, the deflection angle and the magnification
take on a simple form,
\begin{eqnarray}
  \alpha_1={8\pi G\Sigma_0\over \sqrt{2\epsilon}c^2}\tan^{-1}
  \left[{\sqrt{2\epsilon}\cos\phi\over(1-\epsilon\cos2\phi)^{1/2}}
  \right]\;,\nonumber\\
  \alpha_2={8\pi G\Sigma_0\over \sqrt{2\epsilon}c^2}\tanh^{-1}
  \left[{\sqrt{2\epsilon}\sin\phi\over(1-\epsilon\cos2\phi)^{1/2}}
  \right]\;,\nonumber\\
  \mu^{-1}=1-{8\pi G\Sigma_0\over c^2(\theta_1^2+\theta_2^2)^{1/2}
  (1-\epsilon\cos 2\phi)^{1/2}}\;,
\end{eqnarray}
where $\phi$ is the polar angle corresponding to the vector position
$\vec\theta\equiv(\theta_1,\theta_2)$.

Instead of the elliptical density model, it is simpler and often
sufficient to model a galaxy by means of an elliptical effective
lensing potential (Blandford \& Kochanek 1987)
\begin{equation}
  \psi(\theta_1,\theta_2) = \frac{D_{\rm ds}}{D_{\rm s}}\,
  4\pi\frac{\sigma_v^2}{c^2}\,\left[\theta_{\rm c}^2 +
  (1-\epsilon)\theta_1^2 + (1+\epsilon)\theta_2^2\right]^{1/2}\;,
\label{eq:3.29}
\end{equation}
where $\epsilon$ measures the ellipticity. The deflection law and
magnification tensor corresponding to this potential are easily
calculated using the equations given in Sect.\
\ref{NB:subsect3.2}. When $\epsilon$ is large, the elliptical
potential model is inaccurate because it gives rise to dumbbell-shaped
isodensity contours, but for small $\epsilon$, it is a perfectly
viable lens model.

\subsubsection{External Shear}
\label{NB:subsubsect3.5.2}

The environment of a galaxy, including any cluster surrounding the
primary lens, will in general contribute both convergence and
shear. The effective potential due to the local environment then reads
\begin{equation}
  \psi(\theta_1,\theta_2) =
  \frac{\kappa}{2}\,(\theta_1^2+\theta_2^2) +
  \frac{\gamma}{2}\,(\theta_1^2-\theta_2^2)
\label{eq:3.30}
\end{equation}
in the principal axes system of the external shear, where the
convergence $\kappa$ and shear $\gamma$ are locally independent of
$\vec\theta$. An external shear breaks the circular symmetry of a lens
and therefore it often has the same effect as introducing ellipticity
in the lens (Kovner 1987). It is frequently possible to model the same
system either with an elliptical potential or with a circular
potential plus an external shear.

\subsubsection{Image Configurations with a Non-Circularly Symmetric
Lens}
\label{NB:subsubsect3.5.3}

In contrast to the circularly symmetric case, for a non-circular lens
the source, lens and images are not restricted to lie on a line.
Therefore, it is not possible to analyze the problem via sections of
the time delay surface as we did in Figs.\ \ref{NB:fig3.2} and
\ref{NB:fig3.3}. Fermat's principle and the time delay function are
still very useful but it is necessary to visualize the full
two-dimensional surface $t(\vec\theta)$. Those who attended the
lectures in Jerusalem may recall the lecturer demonstrating many of
the qualitative features of imaging by elliptical lenses using a
Mexican hat to simulate the time delay surface. In the following, we
merely state the results.

Figures \ref{NB:fig3.7} and \ref{NB:fig3.8} illustrate the wide
variety of image configurations produced by an elliptical galaxy lens
(or a circularly symmetric lens with external shear). In each panel,
the source plane with caustics is shown on the right, and the image
configurations together with the critical curves are shown on the
left. Compared to the circularly symmetric case, the first notable
difference introduced by ellipticity is that the central caustic which
was point-like is now expanded into a diamond shape; it is referred to
as the {\em astroid\/} caustic (also tangential caustic). Figure
\ref{NB:fig3.7} shows the images of a compact source moving away from
the lens center along a symmetry line (right panel) and a line
bisecting the two symmetry directions (left panel). A source behind
the center of the lens has five images because it is enclosed by two
caustics. One image appears at the lens center, and the four others
form a cross-shaped pattern. When the source is moved outward, two of
the four outer images move toward each other, merge, and disappear as
the source approaches and then crosses the astroid (or tangential)
caustic. Three images remain until the source crosses the radial
caustic, when two more images merge and disappear at the radial
critical curve. A single weakly distorted image is finally left when
the source has crossed the outer caustic. When the source moves toward
a cusp point (right panel of Fig.\ \ref{NB:fig3.7}), three images
merge to form a single image. All the image configurations shown in
Fig.\ \ref{NB:fig3.7} are exhibited by various observed cases of
lensing of QSOs and radio quasars (e.g.\ Keeton \& Kochanek 1996).

\begin{figure}[ht] %PP
%JO \begin{figure}
\begin{center}
\begin{minipage}[t]{0.45\hsize}
\end{minipage}
\begin{minipage}[t]{0.45\hsize}
\end{minipage}
\end{center}
\caption{Compact source moving away from the center of an elliptical
  lens. Left panel: source crossing a fold caustic; right panel:
  source crossing a cusp caustic. Within each panel, the diagram on
  the left shows critical lines and image positions and the diagram on
  the right shows caustics and source positions.}
\label{NB:fig3.7}
\end{figure}

Figure \ref{NB:fig3.8} illustrates what happens when a source with a
larger angular size is imaged by the same lens model as in Fig.\
\ref{NB:fig3.7}. Large arc-like images form which consist either of
three or two merging images, depending on whether the source lies on
top of a cusp in the tangential caustic (top left panel) or on an
inter-cusp segment (a so-called fold caustic, top right panel). If the
source is even larger (bottom panels), four images can merge, giving
rise to complete or incomplete rings. Radio rings such as MG
1131$+$0456 (Hewitt et al.\ 1987) correspond to the configuration
shown at bottom right in Fig.\ \ref{NB:fig3.8}.

\begin{figure}[ht] %PP
%JO \begin{figure}
\begin{center}
\begin{minipage}[t]{0.6\hsize}
\end{minipage}
\end{center}
\caption{Images of resolved sources produced by an elliptical
  lens. Top panels: Large arcs consisting of two or three merging
  images are formed when the source lies on top of a fold section (top
  left panel) or a cusp point (top right panel) of the tangential
  caustic. Bottom panels: A source which covers most of the
  diamond-shaped caustic produces a ring-like image consisting of four
  merging images.}
\label{NB:fig3.8}
\end{figure}

\subsection{Studies of Galaxy Lensing}
\label{NB:subsect3.6}

\subsubsection{Detailed Models of Individual Cases of Lensing}
\label{NB:subsubsect3.6.1}

Gravitational lens observations provide a number of constraints which
can be used to model the mass distribution of the lens. The angular
separation between the images determines the Einstein radius of the
lens and therefore gives the mass $M$ (eq.\ \ref{eq:2.23}) or the
velocity dispersion $\sigma_v$ (eq.\ \ref{eq:3.8}) in simple
models. The appearance or absence of the central image constrains the
core size of the lens. The number of images and their positions
relative to the lens determine the ellipticity of the galaxy, or
equivalently the magnitude and orientation of an external shear. Since
the radial and tangential magnifications of images reflect the local
curvatures of the time-delay surface in the corresponding directions,
the relative image sizes constrain the slope of the density profile of
the lens. This does not work very well if all one has are multiply
imaged point images (Kochanek 1991). However, if the images have radio
structure which can be resolved with VLBI, matters improve
considerably.

\begin{figure}[ht] %PP
%JO \begin{figure}
\begin{center}
\begin{minipage}[t]{0.6\hsize}
\end{minipage}
\end{center}
\caption{Shows an extended source which is mapped into two resolved
  images. While the source and the individual magnification matrices
  ${\cal M}_1$ and ${\cal M}_2$ are not observable, the relative
  magnification matrix ${\cal M}_{12}={\cal M}_1^{-1}{\cal M}_2$ can
  be measured. This matrix provides four independent constraints on
  the lens model.}
\label{NB:fig3.10}
\end{figure}

Figure \ref{NB:fig3.10} shows an extended, irregularly shaped source
which is mapped into two images which are each linear transformations
of the unobservable source. The two transformations are described by
symmetric $2\times2$ magnification matrices ${\cal M}_1$ and ${\cal
M}_2$ (cf.\ eq.\ \ref{eq:3.16}). These matrices cannot be determined
from observations since the original source is not seen. However, the
two images are related to each other by a linear transformation
described by the relative magnification matrix ${\cal M}_{12}={\cal
M}_1^{-1}{\cal M}_2$ which can be measured via VLBI observations
(Gorenstein et al.\ 1988; Falco, Gorenstein, \& Shapiro 1991). The
matrix ${\cal M}_{12}$ is in general not symmetric and thus contains
four independent components, which are each functions of the
parameters of the lens model. In favorable cases, as in QSO
0957$+$561, it is even possible to measure the spatial gradient of
${\cal M}_{12}$ (Garrett et al.\ 1994) which provides additional
constraints on the model.

Radio rings with hundreds of independent pixels are particularly good
for constraining the lens model. As shown in the bottom panels of
Fig.\ \ref{NB:fig3.8}, ring-shaped images are formed from extended
sources which cover a large fraction of the central diamond-shaped
caustic. Rings provide large numbers of independent observational
constraints and are, in principle, capable of providing the most
accurate mass reconstructions of the lens. However, special techniques
are needed for analyzing rings. Three such techniques have been
developed and applied to radio rings, viz.

\begin{itemize}

\item[1] The Ring Cycle algorithm (Kochanek et al.\ 1989) makes use of
the fact that lensing conserves surface brightness. Surface elements
of an extended image which arise from the same source element should
therefore share the same surface brightness (to within observational
errors). This provides a large number of constraints which can be used
to reconstruct the shape of the original source and at the same time
optimize a parameterized lens model.

\item[2] The LensClean technique (Kochanek \& Narayan 1992) is a
generalization of the Ring Cycle algorithm which uses the Clean
algorithm to allow for the finite beam of the radio telescope.

\item[3] LensMEM (Wallington, Narayan, \& Kochanek 1994; Wallington,
Kochanek, \& Narayan 1996) is analogous to LensClean, but uses the
Maximum Entropy Method instead of Clean.

\end{itemize}

\subsubsection{Statistical Modeling of Lens Populations}
\label{NB:subsubsect3.6.2}

The statistics of lensed QSOs can be used to infer statistical
properties of the lens population. In this approach, parameterized
models of the galaxy and QSO populations in the universe are used to
predict the number of lensed QSOs expected to be observed in a given
QSO sample and to model the distributions of various observables such
as the image separation, flux ratio, lens redshift, source redshift,
etc. An important aspect of such studies is the detailed modeling of
selection effects in QSO surveys (Kochanek 1993a) and proper allowance
for magnification bias (Narayan \& Wallington 1993). The lensing
galaxies are usually modeled either as isothermal spheres or in terms
of simple elliptical potentials, with an assumed galaxy luminosity
function and a relation connecting luminosity and galaxy mass (or
velocity dispersion). The QSO number-count as a function of redshift
should be known since it strongly influences the lensing probability.

Statistical studies have been fairly successful in determining
properties of the galaxy population in the universe, especially at
moderate redshifts where direct observations are difficult. Useful
results have been obtained on the number density, velocity
dispersions, core radii, etc. of lenses. Resolved radio QSOs provide
additional information on the internal structure of galaxy lenses such
as their ellipticities (Kochanek 1996b). By and large, the lens
population required to explain the statistics of multiply imaged
optical and radio QSOs turns out to be consistent with the locally
observed galaxy population extrapolated to higher redshifts (Kochanek
1993b; Maoz \& Rix 1993; Surdej et al.\ 1993; see below).

So far, statistical studies of galaxy lensing neglected the
contribution from spirals because their velocity dispersions are
significantly lower than those of ellipticals. However, most of the
lenses found by the CLASS survey (Myers et al.\ 1995) are classified
as S0 or spiral galaxies. This result has recently triggered
investigations of lens models that contain disks in addition to
halos. While realistic disks increase the multiple-image cross
sections of halo-only models only by $\sim10\%$ (Wang \& Turner 1997;
Keeton \& Kochanek 1997), they allow for much more convincing models
of lens systems such as B~1600, where a nearly edge-on disk is
observed (Maller, Flores, \& Primack 1997).

\subsection{Astrophysical Results from Galaxy Lensing}
\label{NB:subsect3.7}

\subsubsection{Galaxy Structure}
\label{NB:subsubsect3.7.1}

The structure of galaxies influences lensing statistics as well as the
appearances of individual lensed objects. Gravitational lensing can
therefore be used to constrain galaxy models in various ways.

As described earlier, galaxy lens models predict a weak central image
whose flux depends on the core radius of the galaxy. The central image
is missing in virtually every known multiply-imaged quasar. The
lensing galaxies in these cases must therefore have very small core
radii, $r_{\rm c}<200$~pc (Wallington \& Narayan 1993; Kassiola \&
Kovner 1993; Grogin \& Narayan 1996).

Kochanek (1993b) has shown that the observed distribution of image
separations in the observed lens sample of quasars requires that most
galaxies must have dark halos with characteristic velocity dispersions
of $\sigma_{\rm dark}^*\sim220\pm20$ km s$^{-1}$. If these dark halos
were absent, virtually no image separations larger than $2''$ would be
produced (Maoz \& Rix 1993), whereas several wide separation examples
are known. Multiply-imaged quasars do not generally constrain the size
of the halo because the constraints only extend out to about the
Einstein radius, which is $\sim10$ kpc at the distance of the
lens. The largest halo inferred from direct modeling of a
multiply-imaged quasar is in the lensing galaxy of QSO 0957$+$561; the
halo of this galaxy has been shown to have a radius of at least
$15h^{-1}$~kpc, where $h=H_0/100~{\rm km\,s^{-1}\,Mpc^{-1}}$ is the
reduced Hubble constant (Grogin \& Narayan 1996). Brainerd et al.\
(1996) investigated weak lensing of background galaxies by foreground
galaxies and found statistical evidence for halos extending out to
radii $\sim100$~kpc. At these radii, they determined that an $L_*$
galaxy must have a mass $\sim10^{12}M_\odot$. Comparable results were
obtained by Dell'Antonio \& Tyson (1996) and Griffiths et al.\ (1996).
Natarajan \& Kneib (1996) show that the sizes of halos around galaxies
in clusters could be inferred by measuring the weak lensing effect of
these galaxies on background sources.

Only in two cases has it been possible to constrain significantly the
radial mass density variation of the lensing galaxy. Assuming a
surface mass density profile $\Sigma\propto r^{-\alpha}$, the best
fitting values of $\alpha$ in these two examples are
\begin{equation}
  \alpha = \left\{
  \begin{array}{lll}
    (0.9-1.1) & \hbox{in MG 1654$+$134} &
    \hbox{(Kochanek 1995a)} \\
    (1.0-1.2) & \hbox{in QSO 0957$+$561} &
    \hbox{(Grogin \& Narayan 1996)} \\
  \end{array}\right.\;.
\label{eq:3.33}
\end{equation}
Both sources have particularly good data --- the first is a radio ring
and the second has extensive VLBI observations --- and it is this
feature that allows a good constraint on $\alpha$. Note that the
density variation is close to isothermal in both cases. Recent
observations of QSO 0957$+$561 with the Hubble Space Telescope
(Bernstein et al.\ 1997) show that the lensing galaxy is shifted by
45~mas from the position assumed by Grogin \& Narayan in their model.
This will modify the estimate of $\alpha$ for this galaxy, but perhaps
by only a fraction of the stated uncertainty.

The observed morphologies of images in lensed quasars are similar to
those shown in Fig.\ \ref{NB:fig3.7}, which means that most lenses are
not circularly symmetric. If the non-circularity is entirely due to
the ellipticity of the galaxy mass, then typical ellipticities are
fairly large, $\sim{\rm E}3-{\rm E}4$ (Kochanek 1996b). However, it is
possible that part of the effect comes from external shear. The data
are currently not able to distinguish very well between the effects of
galaxy ellipticity and external shear. In many well-modeled examples,
the mass ellipticity required to fit the images is larger than the
ellipticity of the galaxy isophotes, suggesting either that the dark
matter is more asymmetric than the luminous matter or that there is a
significant contribution from external shear (Kochanek 1996b; Bar-Kana
1996). Keeton, Kochanek, \& Seljak (1997) find the external cosmic
shear insufficient to explain fully the discrepancy between the
ellipticity of the galaxy isophotes and the ellipticity of the mass
required by lens models. This implies that at least part of the
inferred ellipticity of the mass distribution is intrinsic to the
galaxies.

\subsubsection{Galaxy Formation and Evolution}
\label{NB:subsubsect3.7.2}

The angular separations of multiple images depend on the lens mass,
and the number of observed multiply imaged quasars with a given
separation depends on the number density of galaxies with the
corresponding mass. The usual procedure to set limits on the galaxy
population starts with the present galaxy population and extrapolates
it to higher redshifts assuming some parameterized prescription of
evolution. The parameters are then constrained by comparing the
observed statistics of lensed sources to that predicted by the model
(Kochanek 1993b; Maoz \& Rix 1993; Rix et al.\ 1994; Mao \& Kochanek
1994).

If galaxies formed recently, most of the optical depth for multiple
imaging will be from low-redshift galaxies. An analysis which uses all
the known information on lensed quasars, such as the redshifts of
lenses and sources, the observed fraction of lensed quasars, and the
distribution of image separations, can be used to set limits on how
recently galaxies could have formed. Mao \& Kochanek (1994) conclude
that most galaxies must have collapsed and formed by $z\sim0.8$ if the
universe is well described by the Einstein-de Sitter model.

If elliptical galaxies are assembled from merging spiral galaxies,
then with increasing redshift the present population of ellipticals is
gradually replaced by spirals. This does not affect the probability of
producing lensed quasars as the increase in the number of lens
galaxies at high redshift is compensated by the reduced lensing
cross-sections of these galaxies. However, because of their lower
velocity dispersion, spirals produce smaller image separations than
ellipticals (the image splitting is proportional to $\sigma_v^2$, cf.\
eq.\ \ref{eq:3.8}). Therefore, a merger scenario will predict smaller
image separations in high redshift quasars, and the observed image
separations can be used to constrain the merger rate (Rix et al.\
1994). Assuming that the mass of the galaxies scales with $\sigma_v^4$
and is conserved in mergers, Mao \& Kochanek (1994) find that no
significant mergers could have occurred more recently than $z\sim0.4$
in an Einstein-de Sitter universe.

If the cosmological constant is large, say $\Lambda>0.6$, the volume
per unit redshift is much larger than in an Einstein-de Sitter
universe. For a fixed number density of galaxies, the total number of
available lenses then increases steeply. For such model universes,
lens statistics would be consistent with recent rapid evolution of the
galaxy population. However, studies of gravitational clustering and
structure formation show that galaxies form at high redshifts
precisely when $\Lambda$ is large. When this additional constraint is
included it is found that there is no scenario which allows recent
galaxy formation or evolution in the universe (see also Sect.\
\ref{NB:subsubsect3.7.5}).

Since lensing statistics are fully consistent with the known local
galaxy population extrapolated to redshifts $z\sim1$, the number
densities of any dark ``failed'' galaxies are constrained quite
strongly. As a function of velocity dispersion $\sigma_*$, the current
constraints are (Mao \& Kochanek 1994)
\begin{equation}
  n_{\rm dark} < \left\{
  \begin{array}{lll}
    0.15\,h^3\,{\rm Mpc}^{-3} & \hbox{for} &
      \sigma_*=100\;{\rm km}\;{\rm s}^{-1} \\
    0.032\,h^3\,{\rm Mpc}^{-3} & \hbox{for} &
      \sigma_*=150\;{\rm km}\;{\rm s}^{-1} \\
    0.017\,h^3\,{\rm Mpc}^{-3} & \hbox{for} &
      \sigma_*=200\;{\rm km}\;{\rm s}^{-1} \\
  \end{array}\right.\;.
\label{eq:3.32}
\end{equation}

\subsubsection{Constraint on CDM}
\label{NB:subsubsect3.7.3}

The popular cold dark matter (CDM) scenario of structure formation in
its ``standard'' variant ($\Omega_0=1$, $\Lambda_0=0$ and COBE
normalized) predicts the formation of large numbers of dark matter
halos in the mass range between galaxies and galaxy clusters. The
implications of these halos for lensing were considered by Narayan \&
White (1988) and more recently by Cen et al.\ (1994); Wambsganss et
al.\ (1995); Wambsganss, Cen, \& Ostriker (1996); and Kochanek
(1995b). The latter authors have shown quite convincingly that the
standard CDM model produces many more wide-separation quasar pairs
than observed. For example, a recent search of a subsample of the HST
snapshot survey for multiply imaged QSOs with image separations
between $7''$ and $50''$ found a null result (Maoz et al.\ 1997). To
save CDM, either the normalization of the model needs to be reduced to
$\sigma_8\sim0.5\pm0.2$, or the long-wavelength slope of the power
spectrum needs to be lowered to $n\sim0.5\pm0.2$. Both of these
options are inconsistent with the COBE results. The problem of the
over-production of wide angle pairs is just a manifestation of the
well-known problem that standard COBE-normalized CDM over-produces
cluster-scale mass condensations by a large factor. Models which are
adjusted to fit the observed number density of clusters also satisfy
the gravitational lens constraint.

If the dark halos have large core radii, their central density could
drop below the critical value for lensing and this would reduce the
predicted number of wide-separation lens systems. Large core radii
thus may save standard CDM (Flores \& Primack 1996), but there is some
danger of fine-tuning in such an explanation. As discussed in Sect.\
\ref{NB:subsubsect3.7.1}, galaxy cores are quite small. Therefore, one
needs to invoke a rather abrupt increase of core radius with
increasing halo mass.

\subsubsection{Hubble Constant}
\label{NB:subsubsect3.7.4}

The lens equation is dimensionless, and the positions of images as
well as their magnifications are dimensionless numbers. Therefore,
information on the image configuration alone does not provide any
constraint on the overall scale of the lens geometry or the value of
the Hubble constant. Refsdal (1964) realized that the time delay,
however, is proportional to the absolute scale of the system and does
depend on $H_0$ (cf.\ Fig.\ \ref{NB:fig3.9}).

\begin{figure}[ht] %PP
%JO \begin{figure}
\begin{center}
\begin{minipage}[t]{0.6\hsize}
\end{minipage}
\end{center}
\caption{Sketch of the dependence of the overall scale of a lens
  system on the value of the Hubble constant.}
\label{NB:fig3.9}
\end{figure}

To see this, we first note that the geometrical time delay is simply
proportional to the path lengths of the rays which scale as
$H_0^{-1}$. The potential time delay also scales as $H_0^{-1}$ because
the linear size of the lens and its mass have this scaling.
Therefore, for any gravitational lens system, the quantity
\begin{equation}
  H_0\,\Delta\tau
\label{eq:3.36}
\end{equation}
depends only on the lens model and the geometry of the system. A good
lens model which reproduces the positions and magnifications of the
images provides the scaled time delay $H_0\,\Delta\tau$ between the
images. Therefore, a measurement of the time delay $\Delta\tau$ will
yield the Hubble constant $H_0$ (Refsdal 1964, 1966a).

To measure the time delay, the fluxes of the images need to be
monitored over a period of time significantly longer than the time
delay in order to achieve reasonable accuracy. In fact, the analysis
of the resulting light curves is not straightforward because of uneven
data sampling, and careful and sophisticated data analysis techniques
have to be applied. QSO 0957$+$561 has been monitored both in the
optical (Vanderriest et al.\ 1989; Schild \& Thomson 1993; Kundi\'c et
al.\ 1996) and radio wavebands (Leh\'ar et al.\ 1992; Haarsma et al.\
1996, 1997). Unfortunately, analysis of the data has led to two
claimed time delays:
\begin{equation}
  \Delta\tau = (1.48\pm0.03)\;\hbox{\rm years}
\label{eq:3.37}
\end{equation}
(Press, Rybicki, \& Hewitt 1992a,b) and
\begin{equation}
  \Delta\tau \simeq 1.14\;\hbox{\rm years}
\label{eq:3.38}
\end{equation}
(Schild \& Thomson 1993; Pelt et al.\ 1994, 1996). The discrepancy
appears to have been resolved in favor of the shorter delay. Haarsma
et al.\ (1996) find $\Delta\tau=1.03-1.33\,{\rm years}$, and Kundi\'c
et al.\ (1996) derive $\Delta\tau=417\pm3\,{\rm days}$ using a variety
of statistical techniques.

In addition to a measurement of the time delay, it is also necessary
to develop a reliable model to calculate the value of
$H_0\Delta\tau$. QSO 0957$+$561 has been studied by a number of groups
over the years, with recent work incorporating constraints from VLBI
imaging (Falco et al.\ 1991). Grogin \& Narayan (1996) estimate the
Hubble constant to be given by
\begin{equation}
  H_0 = \left(82\pm6\right)(1-\kappa)
  \left(\frac{\Delta\tau}{1.14~{\rm yr}}\right)^{-1}
  {\rm km\,s^{-1}\,Mpc^{-1}}
\label{eq:3.39}
\end{equation}
where $\kappa$ refers to the unknown convergence due to the cluster
surrounding the lensing galaxy. Since the cluster $\kappa$ cannot be
negative, this result directly gives an upper bound on the Hubble
constant ($H_0 <88~{\rm km\,s^{-1}\,Mpc^{-1}}$ for
$\Delta\tau=1.14$~years). Actually, $\kappa$ can also be modified by
large scale structure along the line of sight. In contrast to the
effect of the cluster, this fluctuation can have either sign, but the
rms amplitude is estimated to be only a few per cent (Seljak 1994;
Bar-Kana 1996). Surpi, Harari, \& Frieman (1996) confirm that
large-scale structure does not modify the functional relationship
between lens observables, and therefore does not affect the
determination of $H_0$.

To obtain an actual value of $H_0$ instead of just an upper bound, we
need an independent estimate of $\kappa$. Studies of weak lensing by
the cluster (Fischer et al.\ 1997) give $\kappa=0.24\pm0.12$
($2\sigma$) at the location of the lens (cf.\ Kundi\'c et al.\ 1996).
This corresponds to $H_0=62_{-13}^{+12}\;{\rm km\,s^{-1}\,Mpc^{-1}}$.
Another technique is to measure the velocity dispersion $\sigma_{\rm
gal}$ of the lensing galaxy, from which it is possible to estimate
$\kappa$ (Falco et al.\ 1992; Grogin \& Narayan 1996). Falco et al.\
(1997) used the Keck telescope to measure $\sigma_{\rm
gal}=279\pm12\;{\rm km\,s^{-1}}$, which corresponds to
$H_0=66\pm7\;{\rm km\,s^{-1}\,Mpc^{-1}}$. Although most models of QSO
0957$+$561 are based on a spherically symmetric galaxy embedded in an
external shear (mostly due to the cluster), introduction of
ellipticity in the galaxy, or a point mass at the galaxy core, or
substructure in the cluster seem to have little effect on the estimate
of $H_0$ (Grogin \& Narayan 1996).

A measurement of the time delay has also been attempted in the
Einstein ring system B~0218$+$ 357. In this case, a single galaxy is
responsible for the small image splitting of $0\farcs3$. The time delay
has been determined to be $12\pm3$ days (1$\sigma$ confidence limit)
which translates to $H_0\sim60\;{\rm km\,s^{-1}\,Mpc^{-1}}$ (Corbett
et al.\ 1996).

Schechter et al.\ (1997) recently announced a time delay of
$\Delta\tau=23.7\pm3.4\,{\rm days}$ between images B and C of the
quadruple lens PG~1115$+$080. Using a different statistical technique,
Bar-Kana (1997) finds $\Delta\tau=25.0^{+3.3}_{-3.8}$ days (95\%
confidence) from the same data. Their best fitting lens model, where
the lens galaxy as well as an associated group of galaxies are modeled
as singular isothermal spheres, gives $H_0=42\pm6\,{\rm km}\,{\rm
s}^{-1}\,{\rm Mpc}^{-1}$. Other models give larger values of $H_0$ but
fit the data less well. Keeton \& Kochanek (1997) have considered a
more general class of models where the lensing galaxy is permitted to
be elliptical, and present a family of models which fit the
PG~1115$+$080 data well. They estimate $H_0=60\pm17\,{\rm km}\,{\rm
s}^{-1}\,{\rm Mpc}^{-1}$. With more accurate data on the position of
the lens galaxy, this estimate could be improved to
$H_0=53^{+10}_{-7}\,{\rm km\,s^{-1}\,Mpc^{-1}}$ (Courbin et al.\
1997).

The determination of $H_0$ through gravitational lensing has a number
of advantages over other techniques.

\begin{itemize}

\item[1] The method works directly with sources at large redshifts,
$z\sim0.5$, whereas most other methods are local (observations within
$\sim100$~Mpc) where peculiar velocities are still comparable to the
Hubble flow.

\item[2] While other determinations of the Hubble constant rely on
distance ladders which progressively reach out to increasing
distances, the measurement via gravitational time delay is a one-shot
procedure. One measures directly the geometrical scale of the lens
system. This means that the lens-based method is absolutely
independent of every other method and at the very least provides a
valuable test of other determinations.

\item[3] The lens-based method is based on fundamental physics (the
theory of light propagation in General Relativity), which is fully
tested in the relevant weak-field limit of gravity. Other methods rely
on models for variable stars (Cepheids) or supernova explosions (Type
II), or empirical calibrations of standard candles (Tully-Fisher
distances, Type I supernovae). The lensing method does require some
information on the ``shapes'' of galaxies which is used to guide the
choice of a parameterized lens model.

\end{itemize}

\subsubsection{Cosmological Constant}
\label{NB:subsubsect3.7.5}

A large cosmological constant $\Lambda_0$ increases the volume per
unit redshift of the universe at high redshift. As Turner (1990)
realized, this means that the relative number of lensed sources for a
fixed comoving number density of galaxies increases rapidly with
increasing $\Lambda_0$. Turning this around it is possible to use the
observed probability of lensing to constrain $\Lambda_0$. This method
has been applied by various authors (Turner 1990; Fukugita \& Turner
1991; Fukugita et al.\ 1992; Maoz \& Rix 1993; Kochanek 1996a), and
the current limit is $\Lambda_0<0.65\quad(2\sigma\;\hbox{confidence
limit})$ for a universe with $\Omega_0+\Lambda_0=1$. With a combined
sample of optical and radio lenses, this limit could be slightly
improved to $\Lambda_0<0.62$ (2$\sigma$; Falco, Kochanek, \& Mu\~noz
1997). Malhotra, Rhoads, \& Turner (1996) claim that there is evidence
for considerable amounts of dust in lensing galaxies. They argue that
the absorption in dusty lenses can reconcile a large cosmological
constant with the observed multiple-image statistics.

A completely independent approach (Kochanek 1992) considers the
redshift distribution of lenses. For a given source redshift, the
probability distribution of $z_{\rm d}$ peaks at higher redshift with
increasing $\Lambda_0$. Once again, by comparing the observations
against the predicted distributions one obtains an upper limit on
$\Lambda_0$. This method is less sensitive than the first, but gives
consistent results.

Another technique consists in comparing the observed QSO image
separations to those expected from the redshifts of lenses and sources
and the magnitudes of the lenses, assuming certain values for
$\Omega_0$ and $\Lambda_0$. The cosmological parameters are then
varied to optimize the agreement with the observations. Applying this
approach to a sample of seven lens systems, Im, Griffiths, \&
Ratnatunga (1997) find $\Lambda_0=0.64^{+0.15}_{-0.26}$ (1$\sigma$
confidence limit) assuming $\Omega_0+\Lambda_0=1$.

\section{Lensing by Galaxy Clusters and Large-Scale Structure}
\label{NB:sect4}

Two distinct types of lensing phenomena are observed with clusters of
galaxies (Fig.\ \ref{NB:fig4.1}):

\begin{itemize}

\item[1] Rich centrally condensed clusters occasionally produce giant
arcs when a background galaxy happens to be aligned with one of the
cluster caustics. These instances of lensing are usually analyzed with
techniques similar to those described in Sect.\ \ref{NB:sect2} for
galaxy lenses. In brief, a parameterized lens model is optimized so as
to obtain a good fit to the observed image.

\item[2] Every cluster produces weakly distorted images of large
numbers of background galaxies. These images are called arclets and
the phenomenon is referred to as weak lensing. With the development of
the Kaiser \& Squires (1993) algorithm and its variants, weak lensing
is being used increasingly to derive parameter-free two-dimensional
mass maps of lensing clusters.

\end{itemize}

\begin{figure}[ht] %PP
%JO \begin{figure}
\begin{center}
\begin{minipage}[t]{0.9\hsize}
\end{minipage}
\end{center}
\caption{{\em Hubble Space Telescope\/} image of the cluster
  Abell~2218, showing a number of arcs and arclets around the two
  centers of the cluster. (NASA HST Archive)}
\label{NB:fig4.1}
\end{figure}

In addition to these two topics, we also discuss in this section weak
lensing by large-scale structure in the universe. This topic promises
to develop into an important branch of gravitational lensing, and
could in principle provide a direct measurement of the primordial
power spectrum $P(k)$ of the density fluctuations in the universe.

\subsection{Strong Lensing by Clusters --- Giant Arcs}
\label{NB:subsect4.1}

\subsubsection{Basic Optics}
\label{NB:subsubsect4.1.1}

We begin by summarizing a few features of generic lenses which we have
already discussed in the previous sections. A lens is fully
characterized by its surface mass density $\Sigma(\vec\theta)$. Strong
lensing, which is accompanied by multiple imaging, requires that the
surface mass density somewhere in the lens should be larger than the
critical surface mass density,
\begin{equation}
  \Sigma\ga\Sigma_{\rm cr} = 0.35\;{\rm g}\,{\rm cm}^{-3}
  \left(\frac{D}{{\rm Gpc}}\right)^{-1}\;,
\label{eq:4.1}
\end{equation}
where $D$ is the effective lensing distance defined in eq.\
(\ref{eq:2.19}). A lens which satisfies this condition produces one or
more caustics. Examples of the caustics produced by an elliptical lens
with a finite core are shown in Fig.\ \ref{NB:fig3.7}. Sources outside
all caustics produce a single image; the number of images increases by
two upon each caustic crossing toward the lens center. As illustrated
in Figs.\ \ref{NB:fig3.7} and \ref{NB:fig3.8}, extended sources like
galaxies produce large arcs if they lie on top of caustics. The
largest arcs are formed from sources on cusp points, because then
three images of a source merge to form the arc (cf.\ the right panel
in Fig.\ \ref{NB:fig3.7} or the top right panel in Fig.\
\ref{NB:fig3.8}). At the so-called ``lips'' and ``beak-to-beak''
caustics, which are related to cusps, similarly large arcs are
formed. Sources on a fold caustic give rise to two rather than three
merging images and thus form moderate arcs.

\subsubsection{Cluster Mass Inside a Giant Arc}
\label{NB:subsubsect4.1.2}

The location of an arc in a cluster provides a simple way to estimate
the projected cluster mass within a circle traced by the arc (cf.\
Fig.\ \ref{NB:fig4.2}). For a circularly symmetric lens, the average
surface mass density $\langle\Sigma\rangle$ within the tangential
critical curve equals the critical surface mass density $\Sigma_{\rm
cr}$. Tangentially oriented large arcs occur approximately at the
tangential critical curves. The radius $\theta_{\rm arc}$ of the
circle traced by the arc therefore gives an estimate of the Einstein
radius $\theta_{\rm E}$ of the cluster.

\begin{figure}[ht] %PP
%JO \begin{figure}
\begin{center}
\begin{minipage}[t]{0.6\hsize}
\end{minipage}
\end{center}
\caption{Tangential arcs constrain the cluster mass within a circle
  traced by the arcs.}
\label{NB:fig4.2}
\end{figure}

Thus we have
\begin{equation}
  \langle\Sigma(\theta_{\rm arc})\rangle \approx
  \langle\Sigma(\theta_{\rm E})\rangle =
\Sigma_{\rm cr}\;,
\label{eq:4.2}
\end{equation}
and we obtain for the mass enclosed by $\theta=\theta_{\rm arc}$
\begin{equation}
  M(\theta) = \Sigma_{\rm cr}\,\pi\,(D_{\rm d}\theta)^2 \approx
  1.1\times10^{14}\,M_\odot\,\left(\frac{\theta}{30''}\right)^2\,
  \left(\frac{D}{1\,{\rm Gpc}}\right)\;.
\label{eq:4.3}
\end{equation}
Assuming an isothermal model for the mass distribution in the cluster
and using eq.\ (\ref{eq:3.8}), we obtain an estimate for the velocity
dispersion of the cluster,
\begin{equation}
  \sigma_v \approx 10^3\,{\rm km\,s}^{-1}\;
  \left(\frac{\theta}{28''}\right)^{1/2}\,
  \left(\frac{D_{\rm s}}{D_{\rm ds}}\right)^{1/2}\;.
\label{eq:4.4}
\end{equation}
In addition to the lensing technique, two other methods are available
to obtain the mass of a cluster: the observed velocity dispersion of
the cluster galaxies can be combined with the virial theorem to obtain
one estimate, and observations of the X-ray gas combined with the
condition of hydrostatic equilibrium provides another. These three
quite independent techniques yield masses which agree with one another
to within a factor $\sim2-3$.

The mass estimate (\ref{eq:4.3}) is based on very simple assumptions.
It can be improved by modeling the arcs with parameterized lens mass
distributions and carrying out more detailed fits of the observed
arcs. We list in Tab.\ \ref{NB:tab4.1} masses, mass-to-blue-light
ratios, and velocity dispersions of three clusters with prominent
arcs. Additional results can be found in the review article by Fort \&
Mellier (1994).

\begin{table}
  \caption{Masses, mass-to-blue-light ratios, and velocity dispersions
  for three clusters with prominent arcs.}
\medskip %PP
  \begin{center}
  \begin{tabular}{|l|cccl|}
  \hline\hline
    Cluster & $M$ & $M/L_B$ & $\sigma$ & Reference \\
    & $(M_\odot)$ & (solar) & km s$^{-1}$ & \\
  \hline\hline
    A~370 & $\sim5\times10^{13}\,h^{-1}$ & $\sim270\,h$ &
    $\sim1350$ &
    Grossman \& Narayan 1989 \\
    &&&& Bergmann et al.\ 1990 \\
    &&&& Kneib et al.\ 1993 \\
  \hline
    A~2390 & $\sim8\times10^{13}\,h^{-1}$ & $\sim240\,h$ &
    $\sim1250$ &
    Pell\'o et al.\ 1991 \\
  \hline
    MS~2137$-$23 & $\sim3\times10^{13}\,h^{-1}$ & $\sim500\,h$ &
    $\sim1100$ &
    Mellier et al.\ 1993 \\
  \hline\hline
  \end{tabular}
  \end{center}
\label{NB:tab4.1}
\end{table}

\subsubsection{Asphericity of Cluster Mass}
\label{NB:subsubsect4.1.3}

The fact that the observed giant arcs never have counter-arcs of
comparable brightness, and rarely have even small counter-arcs,
implies that the lensing geometry has to be non-spherical (Grossman \&
Narayan 1988; Kovner 1989; see also Figs.\ \ref{NB:fig3.7} and
\ref{NB:fig3.8}). Cluster potentials therefore must have substantial
quadrupole and perhaps also higher multipole moments. In the case of
A~370, for example, there are two cD galaxies, and the potential
quadrupole estimated from their separation is consistent with the
quadrupole required to model the observed giant arc (Grossman \&
Narayan 1989). The more detailed model of A~370 by Kneib et al.\
(1993) shows a remarkable agreement between the lensing potential and
the strongly aspheric X-ray emission of the cluster.

Large deviations of the lensing potentials from spherical symmetry
also help increase the probability of producing large arcs. Bergmann
\& Petrosian (1993) argued that the apparent abundance of large arcs
relative to small arcs and arclets can be reconciled with theoretical
expectations if aspheric lens models are taken into
account. Bartelmann \& Weiss (1994) and Bartelmann, Steinmetz, \&
Weiss (1995) showed that the probability for large arcs can be
increased by more than an order of magnitude if aspheric cluster
models with significant substructure are used instead of smooth
spherically symmetric models. The essential reason for this is that
the largest (three-image) arcs are produced by cusp caustics, and
asymmetry increases the number of cusps on the cluster caustics.

\subsubsection{Core Radii}
\label{NB:subsubsect4.1.4}

If a cluster is able to produce large arcs, its surface-mass density
in the core must be approximately supercritical, $\Sigma\ga\Sigma_{\rm
cr}$. If applied to simple lens models, e.g.\ softened isothermal
spheres, this condition requires
\begin{equation}
  \theta_{\rm core} \la 15''\;
  \left(\frac{\sigma_v}{10^3\;{\rm km\,s^{-1}}}\right)^2\;
  \left(\frac{D_{\rm ds}}{D_{\rm s}}\right)\;.
\label{eq:4.5}
\end{equation}
Narayan, Blandford, \& Nityananda (1984) argued that cluster mass
distributions need to have smaller core radii than those derived from
optical and X-ray observations if they are to produce strong
gravitational lens effects. This has been confirmed by many later
efforts to model giant arcs. In virtually every case the core radius
estimated from lensing is significantly smaller than the estimates
from optical and X-ray data. Some representative results on
lens-derived core radii are listed in Tab.\ \ref{NB:tab4.2}, where the
estimates correspond to $H_0=50\;{\rm km\,s^{-1}\,Mpc}^{-1}$.

\begin{table}
  \caption{Limits on cluster core radii from models of large arcs.}
\medskip %PP
  \begin{center}
  \begin{tabular}{|l|cl|}
  \hline\hline
    Cluster & $r_{\rm core}$ & Reference \\
    & (kpc) & \\
  \hline\hline
    A~370 & $<60$ & Grossman \& Narayan (1989) \\
    & $<100$ & Kneib et al.\ (1993) \\
  \hline
    MS~2137$-$23 & $\sim50$ & Mellier et al.\ (1993) \\
  \hline
    Cl~0024$+$1654 & $<130$ & Bonnet  et al.\ (1994) \\
  \hline
    MS~0440$+$0204 & $\ll90$ & Luppino et al.\ (1993) \\
  \hline\hline
  \end{tabular}
  \end{center}
\label{NB:tab4.2}
\end{table}

Statistical analyses based on spherically symmetric cluster models
lead to similar conclusions. Miralda-Escud\'e (1992, 1993) argued that
cluster core radii can hardly be larger than the curvature radii of
large arcs. Wu \& Hammer (1993) claimed that clusters either have to
have singular cores or density profiles much steeper than isothermal
in order to reproduce the abundance of large arcs. Although this
conclusion can substantially be altered once deviations from spherical
symmetry are taken into account (Bartelmann et al.\ 1995), it remains
true that we require $r_{\rm core}\la100$ kpc in all observed
clusters. Cores of this size can also be reconciled with large-arc
statistics.

Interestingly, there are at least two observations which seem to
indicate that cluster cores, although small, must be finite. Fort et
al.\ (1992) discovered a radial arc near the center of MS~2137$-$23,
and Smail et al.\ (1996) found a radial arc in A~370. To produce a
radial arc with a softened isothermal sphere model, the core radius
has to be roughly equal to the distance between the cluster center and
the radial arc (cf.\ Fig.\ \ref{NB:fig3.7}). Mellier, Fort, \& Kneib
(1993) find $r_{\rm core}\ga40$~kpc in MS~2137$-$23, and Smail et al.\
(1996) infer $r_{\rm core}\sim50$~kpc in A~370. Bergmann \& Petrosian
(1993) presented a statistical argument in favor of finite cores by
showing that lens models with singular cores produce fewer large arcs
(relative to small arcs) than observed. The relative abundance
increases with a small finite core. These results, however, have to be
interpreted with caution because it may well be that the softened
isothermal sphere model is inadequate to describe the interiors of
galaxy clusters. While this particular model indeed requires core
radii on the order of the radial critical radius, other lens models
can produce radial arcs without having a flat core, and there are even
singular density profiles which can explain radial arcs
(Miralda-Escud\'e 1995; Bartelmann 1996). Such singular profiles for
the dark matter are consistent with the fairly large core radii
inferred from the X-ray emission of clusters, if the intracluster gas
is isothermal and in hydrostatic equilibrium with the dark-matter
potential (Navarro, Frenk, \& White 1996; J.P. Ostriker, private
communication).

\subsubsection{Radial Density Profile}
\label{NB:subsubsect4.1.5}

Many of the observed giant arcs are unresolved in the radial
direction, some of them even when observed under excellent seeing
conditions or with the {\em Hubble Space Telescope\/}. Since the faint
blue background galaxies which provide the source population for the
arcs seem to be resolved (e.g. Tyson 1995), the giant arcs appear to
be demagnified in width. It was realized by Hammer \& Rigaut (1989)
that spherically symmetric lenses can radially demagnify giant arcs
only if their radial density profiles are steeper than isothermal. The
maximum demagnification is obtained for a point mass lens, where it is
a factor of two. Kovner (1989) and Hammer (1991) demonstrated that,
irrespective of the mass profile and the symmetry of the lens, the
thin dimension of an arc is compressed by a factor
$\approx2(1-\kappa)$, where $\kappa$ is the convergence at the
position of the arc. Arcs which are thinner than the original source
therefore require $\kappa\la0.5$. Since giant arcs have to be located
close to those critical curves in the lens plane along which
$1-\kappa-\gamma=0$, large and thin arcs additionally require
$\gamma\ga0.5$.

In principle, the radius of curvature of large arcs relative to their
distance from the cluster center can be used to constrain the
steepness of the radial density profile (Miralda-Escud\'e 1992, 1993),
but results obtained from observed arcs are not yet conclusive
(Grossman \& Saha 1994). One problem with this method is that
substructure in clusters tends to enlarge curvature radii irrespective
of the mass profile of the dominant component of the cluster
(Miralda-Escud\'e 1993; Bartelmann et al.\ 1995).

Wu \& Hammer (1993) argued for steep mass profiles on statistical
grounds because the observed abundance of large arcs appears to
require highly centrally condensed cluster mass profiles in order to
increase the central mass density of clusters while keeping their
total mass constant. However, their conclusions are based on
spherically symmetric lens models and are significantly changed when
the symmetry assumption is dropped (Bartelmann et al.\ 1995).

It should also be kept in mind that not all arcs are thin. Some
``thick'' and resolved arcs are known (e.g.\ in A~2218,
Pell\'o-Descayre et al.\ 1988; and in A~2390, Pell\'o et al.\ 1991),
and it is quite possible that thin arcs predominate just because they
are more easily detected than thick ones due to observational
selection effects. Also, Miralda-Escud\'e (1992, 1993) has argued that
intrinsic source ellipticity can increase the probability of producing
thin arcs, while Bartelmann et al.\ (1995) showed that the condition
$\kappa\la0.5$ which is required for thin arcs can be more frequently
fulfilled in clusters with substructure where the shear is larger than
in spherically symmetric clusters.

\subsubsection{Mass Sub-Condensations}
\label{NB:subsubsect4.1.6}

The cluster A~370 has two cD galaxies and is a clear example of a
cluster with multiple mass centers. A two-component mass model
centered on the cD galaxies (Kneib et al.\ 1993) fits very well the
lens data as well as X-ray and deep optical images of the
cluster. Abell~2390 is an interesting example because it contains a
``straight arc'' (Pell\'o et al.\ 1991, see also Mathez et al.\ 1992)
which can be produced only with either a lips or a beak-to-beak
caustic (Kassiola, Kovner, \& Blandford 1992). If the arc is modeled
with a lips caustic, it requires the mass peak to be close to the
location of the arc, but this is not where the cluster light is
centered. With a beak-to-beak caustic, the model requires two separate
mass condensations, one of which could be at the peak of the
luminosity, but then the other has to be a dark condensation. Pierre
et al.\ (1996) find enhanced X-ray emission at a plausible position of
the secondary mass clump, and from a weak lensing analysis Squires et
al.\ (1996b) find a mass map which is consistent with a mass
condensation at the location of the enhanced X-ray emission.

Abell~370 and A~2390 are the most obvious examples of what is probably
a widespread phenomenon, namely that clusters are in general not fully
relaxed but have substructure as a result of ongoing evolution. If
clusters are frequently clumpy, this can lead to systematic effects in
the statistics of arcs and in the derived cluster parameters
(Bartelmann et al.\ 1995; Bartelmann 1995b).

\subsubsection{Lensing Results vs. Other Mass Determinations}
\label{NB:subsubsect4.1.7}

\paragraph{Enclosed Mass}

Three different methods are currently used to estimate cluster
masses. Galaxy velocity dispersions yield a mass estimate from the
virial theorem, and hence the galaxies have to be in virial
equilibrium for such estimates to be valid. It may also be that the
velocities of cluster galaxies are biased relative to the velocities
of the dark matter particles (Carlberg 1994), though current estimates
suggest that the bias is no more than about 10\%. The X-ray emission
of rich galaxy clusters is dominated by free-free emission of thermal
electrons and therefore depends on the squared density of the
intracluster gas, which in turn traces the gravitational potential of
the clusters. Such estimates usually assume that the cluster gas is in
thermal hydrostatic equilibrium and that the potential is at least
approximately spherically symmetric. Finally, large arcs in clusters
provide a mass estimate through eq.\ (\ref{eq:4.3}) or by more
detailed modeling. These three mass estimates are in qualitative
agreement with each other up to factors of $\approx2-3$.

Miralda-Escud\'e \& Babul (1995) compared X-ray and large-arc mass
estimates for the clusters A~1689, A~2163 and A~2218. They took into
account deviations from spherical symmetry and obtained lensing masses
from individual lens models which reproduce the observed arcs. They
arrived at the conclusion that in A~1689 and A~2218 the mass required
for producing the large arcs is higher by a factor of $2-2.5$ than the
mass required for the X-ray emission, and proposed a variety of
reasons for such a discrepancy, among them projection effects and
non-thermal pressure support. Loeb \& Mao (1994) specifically
suggested that strong turbulence and magnetic fields in the
intracluster gas may constitute a significant non-thermal pressure
component in A~2218 and thus render the X-ray mass estimate too
low. Bartelmann \& Steinmetz (1996) used gas dynamical cluster
simulations to compare their X-ray and lensing properties. They found
a similar discrepancy as that identified by Miralda-Escud\'e \& Babul
(1995) in those clusters that show structure in the distribution of
line-of-sight velocities of the cluster particles, indicative of
merging or infall along the line-of-sight. The discrepancy is probably
due to projection effects.

Bartelmann (1995b) showed that cluster mass estimates obtained from
large arcs by straightforward application of eq.\ (\ref{eq:4.3}) are
systematically too high by a factor of $\approx1.6$ on average, and by
as much as a factor of $\approx2$ in 1 out of 5 cases. This
discrepancy arises because eq.\ (\ref{eq:4.3}) assumes a smooth
spherically symmetric mass distribution whereas realistic clusters are
asymmetric and have substructure. Note that Daines et al.\ (1996)
found evidence for two or more mass condensations along the line of
sight toward A~1689, while the arclets in A~2218 show at least two
mass concentrations. It appears that cluster mass estimates from
lensing require detailed lens models in order to be accurate to better
than $\approx30-50$ per cent. In the case of MS~1224, Fahlman et al.\
(1994) and Carlberg, Yee, \& Ellingson (1994) have obtained masses
using the Kaiser \& Squires weak-lensing cluster reconstruction
method. Their mass estimates are $2-3$ times higher than the cluster's
virial mass. Carlberg et al.\ find evidence from velocity measurements
that there is a second poor cluster in the foreground of MS~1224 which
may explain the result. All of these mass discrepancies illustrate
that cluster masses must still be considered uncertain to a factor of
$\approx2$ in general.

\paragraph{Core Radii}

Lensing estimates of cluster core radii are generally much smaller
than the core radii obtained from optical or X-ray data. The upper
limits on the core radii from lensing are fairly robust and probably
reliable. Many clusters with large arcs have cD galaxies which can
steepen the central mass profile of the cluster. However, there are
also non-cD clusters with giant arcs, e.g.\ A~1689 and Cl~1409 (Tyson
1990), and MS~0440$+$02 (Luppino et al.\ 1993). In fact, Tyson (1990)
claims that these two clusters have cores smaller than
$100\,h^{-1}$~kpc, similar to upper limits for core radii found in
other arc clusters with cD galaxies. As mentioned in Sect.\
\ref{NB:subsubsect4.1.4}, even the occurrence of radial arcs in
clusters does not necessarily require a non-singular core, and so all
the lensing data are consistent with singular cores in clusters. The
X-ray core radii depend on whether or not the cooling regions of
clusters are included in the emissivity profile fits, because the
cooling radii are of the same order of magnitude as the core radii. If
cooling is included, the best-fit core radii are reduced by a factor
of $\simeq4$ (Gerbal et al.\ 1992; Durret et al.\ 1994). Also,
isothermal gas in hydrostatic equilibrium in a singular dark-matter
distribution develops a flat core with a core radius similar to those
observed. Therefore, the strongly peaked mass distributions required
for lensing seem to be quite compatible with the extended X-ray cores
observed.

\paragraph{Does Mass Follow Light?}

Leaving the core radius aside, does mass follow light? It is clear
that the mass cannot be as concentrated within the galaxies as the
optical light is (e.g.\ Bergmann, Petrosian, \& Lynds 1990). However,
if the optical light is smoothed and assumed to trace the mass, then
the resulting mass distribution is probably not very different from
the true mass distribution. For instance, in A~370, the elongation of
the mass distribution required for the giant arc is along the line
connecting the two cD galaxies in the cluster (Grossman \& Narayan
1989) and in fact Kneib et al.\ (1993) are able to achieve an
excellent fit of the giant arc and several arclets with two mass
concentrations surrounding the two cDs. Their model potential also
agrees very well with the X-ray emission of the cluster. In MS~2137,
the optical halo is elongated in the direction indicated by the arcs
for the overall mass asymmetry (Mellier et al.\ 1993), and in Cl~0024,
the mass distribution is elongated in the same direction as the galaxy
distribution (Wallington, Kochanek, \& Koo 1995). Smail et al.\ (1995)
find that the mass maps of two clusters reconstructed from weak
lensing agree fairly well with their X-ray emission. An important
counterexample is the cluster A~2390, where the straight arc requires
a mass concentration which is completely dark in the optical (Kassiola
et al.\ 1992). Pierre et al.\ (1996), however, find excess X-ray
emission at a position compatible with the arc.

\paragraph{What Kinds of Clusters Produce Giant Arcs?}

Which parameters determine whether or not a galaxy cluster is able to
produce large arcs? Clearly, large velocity dispersions and small core
radii favor the formation of arcs. As argued earlier, intrinsic
asymmetries and substructure also increase the ability of clusters to
produce arcs because they increase the shear and the number of cusps
in the caustics.

The abundance of arcs in X-ray luminous clusters appears to be higher
than in optically selected clusters. At least a quarter, maybe half,
of the 38 X-ray bright clusters selected by Le F\`evre et al.\ (1994)
contain large arcs, while Smail et al.\ (1991) found only one large
arc in a sample of 19 distant optically selected clusters. However,
some clusters which are prominent lenses (A~370, A~1689, A~2218) are
moderate X-ray sources, while other clusters which are very luminous
X-ray sources (A~2163, Cl~1455) are poor lenses. The correlation
between X-ray brightness and enhanced occurrence of arcs may suggest
that X-ray bright clusters are more massive and/or more centrally
condensed than X-ray quiet clusters.

Substructure appears to be at least as important as X-ray brightness
for producing giant arcs. For example, A~370, A~1689 and A~2218 all
seem to have clumpy mass distributions. Bartelmann \& Steinmetz (1996)
used numerical cluster simulations to show that the optical depth for
arc formation is dominated by clusters with intermediate rather than
the highest X-ray luminosities.

Another possibility is that giant arcs preferentially form in clusters
with cD galaxies. A~370, for instance, even has two cDs. However,
non-cD clusters with giant arcs are known, e.g.\ A~1689, Cl~1409
(Tyson et al.\ 1990), and MS~0440$+$02 (Luppino et al.\ 1993).

\subsection{Weak Lensing by Clusters --- Arclets}
\label{NB:subsect4.2}

In addition to the occasional giant arc, which is produced when a
source happens to straddle a caustic, a lensing cluster also produces
a large number of weakly distorted images of other background sources
which are not located near caustics. These are the arclets. There is a
population of distant blue galaxies in the universe whose spatial
density reaches $50-100$ galaxies per square arc minute at faint
magnitudes (Tyson 1988). Each cluster therefore has on the order of
$50-100$ arclets per square arc minute exhibiting a coherent pattern
of distortions. Arclets were first detected by Fort et al.\ (1988).

The separations between arclets, typically $\sim(5-10)''$, are much
smaller than the scale over which the gravitational potential of a
cluster as a whole changes appreciably. The weak and noisy signals
from several individual arclets can therefore be averaged by
statistical techniques to get an idea of the mass distribution of a
cluster. This technique was first demonstrated by Tyson, Valdes, \&
Wenk (1990). Kochanek (1990) and Miralda-Escud\'e (1991a) studied how
parameterized cluster lens models can be constrained with arclet data.

The first systematic and parameter-free procedure to convert the
observed ellipticities of arclet images to a surface density map
$\Sigma(\vec\theta)$ of the lensing cluster was developed by Kaiser \&
Squires (1993). An ambiguity intrinsic to all such inversion methods
which are based on shear information alone was identified by Seitz \&
Schneider (1995a). This ambiguity can be resolved by including
information on the convergence of the cluster; methods for this were
developed by Broadhurst, Taylor, \& Peacock (1995) and Bartelmann \&
Narayan (1995a).

\subsubsection{The Kaiser \& Squires Algorithm}
\label{NB:subsubsect4.2.1}

The technique of Kaiser \& Squires (1993) is based on the fact that
both convergence $\kappa(\vec\theta)$ and shear
$\gamma_{1,2}(\vec\theta)$ are linear combinations of second
derivatives of the effective lensing potential
$\psi(\vec\theta)$. There is thus a mathematical relation connecting
the two. In the Kaiser \& Squires method one first estimates
$\gamma_{1,2} (\vec\theta)$ by measuring the weak distortions of
background galaxy images, and then uses the relation to infer
$\kappa(\vec\theta)$. The surface density of the lens is then obtained
from $\Sigma (\vec\theta)=\Sigma_{\rm cr}\kappa(\vec\theta)$ (see eq.\
\ref{eq:3.13}).

As shown in Sect.\ \ref{NB:sect3}, $\kappa$ and $\gamma_{1,2}$ are
given by
\begin{eqnarray}
  \kappa(\vec\theta) &=& \frac{1}{2}\,\left(
  \frac{\partial^2\psi(\vec\theta)}{\partial\theta_1^2} +
  \frac{\partial^2\psi(\vec\theta)}{\partial\theta_2^2}
  \right)\;,\nonumber\\
  \gamma_1(\vec\theta) &=& \frac{1}{2}\,\left(
  \frac{\partial^2\psi(\vec\theta)}{\partial\theta_1^2} -
  \frac{\partial^2\psi(\vec\theta)}{\partial\theta_2^2}
  \right)\;,\nonumber\\
  \gamma_2(\vec\theta) &=& \frac{\partial^2\psi(\vec\theta)}
  {\partial\theta_1\partial\theta_2}\;.\nonumber\\
\label{eq:4.6}
\end{eqnarray}
If we introduce Fourier transforms of $\kappa$, $\gamma_{1,2}$, and
$\psi$ (which we denote by hats on the symbols), we have
\begin{eqnarray}
  \hat\kappa(\vec k) &=&
  -\frac{1}{2}\,(k_1^2+k_2^2)\hat\psi(\vec k)\;,\nonumber\\
  \hat\gamma_1(\vec k) &=&
  -\frac{1}{2}\,(k_1^2-k_2^2)\hat\psi(\vec k)\;,\nonumber\\
  \hat\gamma_2(\vec k) &=&
  -k_1k_2\hat\psi(\vec k)\;,\nonumber\\
\label{eq:4.7}
\end{eqnarray}
where $\vec k$ is the two dimensional wave vector conjugate to
$\vec\theta$. The relation between $\kappa$ and $\gamma_{1,2}$ in
Fourier space can then be written
\begin{eqnarray}
  \pmatrix{\hat\gamma_1 \cr \hat\gamma_2 \cr} &=&
  k^{-2}\,\pmatrix{(k_1^2-k_2^2) \cr 2k_1k_2 \cr}\,\hat\kappa\;,
  \nonumber\\
  \hat\kappa &=& k^{-2}\,\left[(k_1^2-k_2^2),(2k_1k_2)\right]\,
  \pmatrix{\hat\gamma_1 \cr \hat\gamma_2 \cr}\;.\nonumber\\
\label{eq:4.8}
\end{eqnarray}
If the shear components $\gamma_{1,2}(\vec\theta)$ have been measured,
we can solve for $\hat\kappa(\vec k)$ in Fourier space, and this can
be back transformed to obtain $\kappa(\vec\theta)$ and thereby
$\Sigma(\vec\theta)$. Equivalently, we can write the relationship as a
convolution in $\vec\theta$ space,
\begin{equation}
  \kappa(\vec\theta) = \frac{1}{\pi}\,\int\,d^2\theta'\,
  {\rm Re}\left[
    {\cal D}^*(\vec\theta-\vec\theta')\,\gamma(\vec\theta')
  \right]\;,
\label{eq:4.9}
\end{equation}
where ${\cal D}$ is the complex convolution kernel,
\begin{equation}
  {\cal D}(\vec\theta) =
  \frac{(\theta_2^2-\theta_1^2) - 2{\rm i}\theta_1\theta_2}
  {\theta^4}\;,
\label{eq:4.10}
\end{equation}
and $\gamma(\vec\theta)$ is the complex shear,
\begin{equation}
  \gamma(\vec\theta) =
  \gamma_1(\vec\theta) + {\rm i}\gamma_2(\vec\theta)\;.
\label{eq:4.11}
\end{equation}
The asterisk denotes complex conjugation.

The key to the Kaiser \& Squires method is that the shear field
$\gamma(\vec\theta)$ can be measured. (Elaborate techniques to do so
were described by Bonnet \& Mellier 1995 and Kaiser, Squires, \&
Broadhurst 1995.) If we define the ellipticity of an image as
\begin{equation}
  \epsilon = \epsilon_1 + {\rm i}\epsilon_2 =
  \frac{1-r}{1+r}\,{\rm e}^{2{\rm i}\phi}\;,
  \quad r\equiv\frac{b}{a}\;,
\label{eq:4.12}
\end{equation}
where $\phi$ is the position angle of the ellipse and $a$ and $b$ are
its major and minor axes, respectively, we see from eq.\
(\ref{eq:3.22}) that the average ellipticity induced by lensing is
\begin{equation}
  \langle\epsilon\rangle =
  \left\langle\frac{\gamma}{1-\kappa}\right\rangle\;,
\label{eq:4.13}
\end{equation}
where the angle brackets refer to averages over a finite area of the
sky. In the limit of weak lensing, $\kappa\ll1$ and $|\gamma|\ll1$,
and the mean ellipticity directly gives the shear,
\begin{equation}
  \left\langle\gamma_1(\vec\theta)\right\rangle
     \approx \left\langle\epsilon_1(\vec\theta)\right\rangle
  \;,\quad
  \left\langle\gamma_2(\vec\theta)\right\rangle
     \approx \left\langle\epsilon_2(\vec\theta)\right\rangle
  \;.
\label{eq:4.14}
\end{equation}
The $\gamma_{1}(\vec\theta)$, $\gamma_{2}(\vec\theta)$ fields so
obtained can be transformed using the integral (\ref{eq:4.9}) to
obtain $\kappa(\vec\theta)$ and thereby $\Sigma(\vec\theta)$. The
quantities $\langle\epsilon_{1}(\vec\theta)\rangle$ and
$\langle\epsilon_{2}(\vec\theta)\rangle$ in (\ref{eq:4.14}) have to be
obtained by averaging over sufficient numbers of weakly lensed sources
to have a reasonable signal-to-noise ratio.

\subsubsection{Practical Details and Subtleties}
\label{NB:subsubsect4.2.2}

In practice, several difficulties complicate the application of the
elegant inversion technique summarized by eq.\
(\ref{eq:4.9}). Atmospheric turbulence causes images taken by
ground-based telescopes to be blurred. As a result, elliptical images
tend to be circularized so that ground-based telescopes measure a
lower limit to the actual shear signal. This difficulty is not present
for space-based observations.

The point-spread function of the telescope can be anisotropic and can
vary across the observed field. An intrinsically circular image can
therefore be imaged as an ellipse just because of astigmatism of the
telescope. Subtle effects like slight tracking errors of the telescope
or wind at the telescope site can also introduce a spurious shear
signal.

In principle, all these effects can be corrected for. Given the seeing
and the intrinsic brightness distribution of the image, the amount of
circularization due to seeing can be estimated and taken into
account. The shape of the point-spread function and its variation
across the image plane of the telescope can also be
calibrated. However, since the shear signal especially in the
outskirts of a cluster is weak, the effects have to be determined with
high precision, and this is a challenge.

The need to average over several background galaxy images introduces a
resolution limit to the cluster reconstruction. Assuming 50 galaxies
per square arc minute, the typical separation of two galaxies is
$\sim8''$. If the average is taken over $\sim10$ galaxies, the spatial
resolution is limited to $\sim30''$.

We have seen in eq.\ (\ref{eq:4.13}) that the observed ellipticities
strictly do not measure $\gamma$, but rather a combination of $\kappa$
and $\gamma$,
\begin{equation}
  \langle\epsilon\rangle = \langle g\rangle \equiv
  \left\langle\frac{\gamma}{1-\kappa}\right\rangle\;.
\label{eq:4.15}
\end{equation}
Inserting $\gamma=\langle\epsilon\rangle(1-\kappa)$ into the
reconstruction equation (\ref{eq:4.9}) yields an integral equation for
$\kappa$ which can be solved iteratively. This procedure, however,
reveals a weakness of the method. Any reconstruction technique which
is based on measurements of image ellipticities alone is insensitive
to isotropic expansions of the images. The measured ellipticities are
thus invariant against replacing the Jacobian matrix ${\cal A}$ by
some scalar multiple $\lambda\,{\cal A}$ of it. Putting
\begin{equation}
  {\cal A}' = \lambda\,{\cal A} =
  \lambda\,\pmatrix{1-\kappa-\gamma_1 & -\gamma_2 \cr
  -\gamma_2 & 1-\kappa+\gamma_1 \cr}\;,
\label{eq:4.16}
\end{equation}
we see that scaling ${\cal A}$ with $\lambda$ is equivalent to the
following transformations of $\kappa$ and $\gamma$,
\begin{equation}
  1-\kappa' = \lambda\,(1-\kappa)\;,\quad
  \gamma' = \lambda\,\gamma\;.
\label{eq:4.17}
\end{equation}
Manifestly, this transformation leaves $g$ invariant. We thus have a
one-parameter ambiguity in shear-based reconstruction techniques,
\begin{equation}
  \kappa \to \lambda\kappa + (1-\lambda)\;,
\label{eq:4.18}
\end{equation}
with $\lambda$ an arbitrary scalar constant.

This invariance transformation was highlighted by Schneider \& Seitz
(1995) and was originally discovered by Falco, Gorenstein, \& Shapiro
(1985) in the context of galaxy lensing. If $\lambda\la1$, the
transformation is equivalent to replacing $\kappa$ by $\kappa$ plus a
sheet of constant surface mass density $1-\lambda$. The transformation
(\ref{eq:4.18}) is therefore referred to as the mass-sheet degeneracy.

Another weakness of the Kaiser \& Squires method is that the
reconstruction equation (\ref{eq:4.9}) requires a convolution to be
performed over the entire $\vec\theta$ plane. Observational data
however are available only over a finite field. Ignoring everything
outside the field and restricting the range of integration to the
actual field is equivalent to setting $\gamma=0$ outside the
field. For circularly symmetric mass distributions, this implies
vanishing total mass within the field. The influence of the finiteness
of the field can therefore be quite severe.

\begin{figure}[ht] %PP
%JO \begin{figure}
\begin{center}
\begin{minipage}[t]{0.45\hsize}
\end{minipage}
\begin{minipage}[t]{0.45\hsize}
\end{minipage}
\end{center}
\caption{{\em HST\/} image of the cluster Cl~0024, overlaid on the
  left with the shear field obtained from an observation of arclets
  with the {\em Canada-France Hawaii Telescope\/} (Y.\ Mellier \& B.\
  Fort), and on the right with the reconstructed surface-mass density
  determined from the shear field (C.\ Seitz et al.). The
  reconstruction was done with a non-linear, finite-field algorithm.}
\label{NB:fig4.3}
\end{figure}

Finally, the reconstruction yields $\kappa(\vec\theta)$, and in order
to calculate the surface mass density $\Sigma(\vec\theta)$ we must
know the critical density $\Sigma_{\rm cr}$, but since we do not know
the redshifts of the sources there is a scaling uncertainty in this
quantity. For a lens with given surface mass density, the distortion
increases with increasing source redshift. If the sources are at much
higher redshifts than the cluster, the influence of the source
redshift becomes weak. Therefore, this uncertainty is less serious for
low redshift clusters.

Nearly all the problems mentioned above have been addressed and
solved. The solutions are discussed in the following subsections.

\subsubsection{Eliminating the Mass Sheet Degeneracy by Measuring the
Convergence}
\label{NB:subsubsect4.2.3}

By eq.\ (\ref{eq:3.23}),
\begin{equation}
  \mu = \frac{1}{(1-\kappa)^2-\gamma^2}\;,
\label{eq:4.19}
\end{equation}
and so the magnification scales with $\lambda$ as
$\mu\propto\lambda^{-2}$. Therefore, the mass-sheet degeneracy can be
broken by measuring the magnification $\mu$ of the images in addition
to the shear (Broadhurst et al.\ 1995). Two methods have been proposed
to measure $\mu$. The first relies on comparing the galaxy counts in
the cluster field with those in an unlensed ``empty'' field
(Broadhurst et al.\ 1995). The observed counts of galaxies brighter
than some limiting magnitude $m$ are related to the intrinsic counts
through
\begin{equation}
  N'(m) = N_0(m)\,\mu^{2.5\,s-1}\;,
\label{eq:4.20}
\end{equation}
where $s$ is the logarithmic slope of the intrinsic number count
function,
\begin{equation}
  s = \frac{d\log N(m)}{dm}\;.
\label{eq:4.21}
\end{equation}
In blue light, $s\sim0.4$, and thus $N'(m)\sim N(m)$ independent of
the magnification, but in red light $s\sim0.15$, and the magnification
leads to a dilution of galaxies behind clusters. The reduction of red
galaxy counts behind the cluster A~1689 has been detected by
Broadhurst (1995).

The other method is to compare the {\em sizes\/} of galaxies in the
cluster field to those of similar galaxies in empty fields. Since
lensing conserves surface brightness, it is most convenient to match
galaxies with equal surface brightness while making this comparison
(Bartelmann \& Narayan 1995a). The magnification is then simply the
ratio between the sizes of lensed and unlensed galaxies. Labeling
galaxies by their surface brightness has the further advantage that
the surface brightness is a steep function of galaxy redshift, which
allows the user to probe the change of lens efficiency with source
redshift (see below).

\subsubsection{Determining Source Redshifts}
\label{NB:subsubsect4.2.4}

For a given cluster, the strength of distortion and magnification due
to lensing increases with increasing source redshift $z_{\rm s}$. The
mean redshift $\bar z_{\rm s}$ of sources as a function of apparent
magnitude $m$ can thus be inferred by studying the mean strength of
the lensing signal vs.\ $m$ (Kaiser 1995; Kneib et al.\ 1996).

The surface brightness $S$ probably provides a better label for
galaxies than the apparent magnitude because it depends steeply on
redshift and is unchanged by lensing. Bartelmann \& Narayan (1995a)
have developed an algorithm, which they named the lens parallax
method, to reconstruct the cluster mass distributions and to infer
simultaneously $\bar z_{\rm s}$ as a function of the surface
brightness. In simulations, data from $\sim10$ cluster fields and an
equal number of empty comparison fields were sufficient to determine
the cluster masses to $\sim\pm5\%$ and the galaxy redshifts to
$\sim\pm10\%$ accuracy. The inclusion of galaxy sizes in the iterative
lens-parallax algorithm breaks the mass-sheet degeneracy, thereby
removing the ambiguities in shear-based cluster reconstruction
techniques arising from the transformation (\ref{eq:4.18}) and from
the unknown redshift distribution of the sources.

\subsubsection{Finite Field Methods}
\label{NB:subsubsect4.2.5}

As emphasized previously, the inversion equation (\ref{eq:4.9})
requires a convolution to be performed over the entire real plane. The
fact that data are always restricted to a finite field thus introduces
a severe bias in the reconstruction. Modified reconstruction kernels
have been suggested to overcome this limitation.

Consider the relation (Kaiser 1995)
\begin{equation}
  \vec\nabla\kappa = \pmatrix{
  \gamma_{1,1} + \gamma_{2,2} \cr
  \gamma_{2,1} - \gamma_{1,2} \cr}\;.
\label{eq:4.22}
\end{equation}
This shows that the convergence at any point $\vec\theta$ in the data
field is related by a line integral to the convergence at another
point $\vec\theta_0$,
\begin{equation}
  \kappa(\vec\theta) = \kappa(\vec\theta_0) +
  \int_{\vec\theta_0}^{\vec\theta}d{\vec l}\cdot
  \vec\nabla\kappa[\vec\theta(\vec l)]\;.
\label{eq:4.23}
\end{equation}
If the starting point $\vec\theta_0$ is far from the cluster center,
$\kappa(\vec\theta_0)$ may be expected to be small and can be
neglected. For each starting point $\vec\theta_0$, eq.\
(\ref{eq:4.23}) yields an estimate for
$\kappa(\vec\theta)-\kappa(\vec\theta_0)$, and by averaging over all
chosen $\vec\theta_0$ modified reconstruction kernels can be
constructed (Schneider 1995; Kaiser et al.\ 1995; Bartelmann 1995c;
Seitz \& Schneider 1996). Various choices for the set of starting
positions $\vec\theta_0$ have been suggested. For instance, one can
divide the observed field into an inner region centered on the cluster
and take as $\vec\theta_0$ all points in the rest of the
field. Another possibility is to take $\vec\theta_0$ from the entire
field. In both cases, the result is $\kappa(\vec\theta)-\bar\kappa$,
where $\bar\kappa$ is the average convergence in the region from which
the points $\vec\theta_0$ were chosen. The average $\bar\kappa$ is
unknown, of course, and thus a reconstruction based on eq.\
(\ref{eq:4.23}) yields $\kappa$ only up to a constant. Equation
(\ref{eq:4.23}) therefore explicitly displays the mass sheet
degeneracy since the final answer depends on the choice of the unknown
$\kappa(\vec\theta_0)$.

A different approach (Bartelmann et al.\ 1996) employs the fact that
$\kappa$ and $\gamma$ are linear combinations of second derivatives of
the same effective lensing potential $\psi$. In this method one
reconstructs $\psi$ rather than $\kappa$. If both $\kappa$ and
$\gamma$ can be measured through image distortions and magnifications
(with different accuracies), then a straightforward finite-field
Maximum-Likelihood can be developed to construct $\psi(\vec\theta)$ on
a finite grid such that it optimally reproduces the observed
magnifications and distortions. It is easy in this approach to
incorporate measurement accuracies, correlations in the data,
selection effects etc.\ to achieve an optimal result.

\subsubsection{Results from Weak Lensing}
\label{NB:subsubsect4.2.6}

The cluster reconstruction technique of Kaiser \& Squires and variants
thereof have been applied to a number of clusters and several more are
being analyzed. We summarize some results in Tab.\ \ref{NB:tab4.3},
focusing on the mass-to-light ratios of clusters and the degree of
agreement between weak lensing and other independent studies of the
same clusters.

\begin{table}
  \caption{Mass-to-light ratios of several clusters derived from weak
  lensing.}
\medskip %PP
  \begin{center}
  \begin{tabular}{|l|cll|}
  \hline\hline
    Cluster & $M/L$ & Remark & Reference \\
  \hline\hline
    MS~1224 & 8$00\,h$ & virial mass $\sim3$ times smaller &
    Fahlman et al.\ \\
    && ($\sigma_v=770$ km s$^{-1}$) & (1994) \\
    && reconstruction out to $\sim3'$ & \\
  \hline
    A~1689 & $(400\pm60)\,h$ & mass smoother than light &
    Tyson \& Fischer \\
    && near center; mass steeper & (1995) \\
    && than isothermal from & Kaiser (1995) \\
    && $(200-1000)\,h^{-1}$ kpc & \\
  \hline
    Cl~1455 & $520\,h$ & dark matter more concen- &
    Smail et al.\ \\
    && trated than galaxies & (1995) \\
  \hline
    Cl~0016 & $740\,h$ & dark matter more concen- &
    Smail et al.\ \\
    && trated than galaxies & (1995) \\
  \hline
    A~2218 & $440\,h$ & gas mass fraction &
    Squires et al.\ \\
    && $<4\%\,h^{-3/2}$ & (1996a) \\
  \hline
    A~851 & $200\,h$ & mass distribution agrees &
    Seitz et al.\ (1996) \\
    && with galaxies and X-rays & \\
  \hline
    A~2163 & $(300\pm100)\,h$ & gas mass fraction &
    Squires et al.\ \\
    && $\sim7\%\,h^{-3/2}$ & (1997) \\
  \hline\hline
  \end{tabular}
  \end{center}
\label{NB:tab4.3}
\end{table}

Mass-to-light ratios inferred from weak lensing are generally quite
high, $\sim400\,h$ in solar units (cf.\ table \ref{NB:tab4.3} and,
e.g., Smail et al.\ 1997). The recent detection of a significant
shear signal in the cluster MS~1054$-$03 at redshift $0.83$ (Luppino
\& Kaiser 1997) indicates that the source galaxies either are at very
high redshifts, $z\ga(2-3)$, or that the mass-to-light ratio in this
cluster is exceptionally high; if the galaxy redshifts are $z\la1$,
the mass-to-light ratio needs to be $\ga1600\,h$.

The measurement of a coherent weak shear pattern out to a distance of
almost $1.5$ Mpc from the center of the cluster Cl~0024$+$1654 by
Bonnet, Mellier, \& Fort (1994) demonstrates a promising method of
constraining cluster mass profiles. These observations show that the
density decreases rapidly outward, though the data are compatible both
with an isothermal profile and a steeper de Vaucouleurs profile. Tyson
\& Fischer (1995) find the mass profile in A~1689 to be steeper than
isothermal. Squires et al.\ (1996b) derived the mass profile in A~2390
and showed that it is compatible with both an isothermal profile and
steeper profiles. Quite generally, the weak-lensing results on
clusters indicate that the smoothed light distribution follows the
mass well. Moreover, mass estimates from weak lensing and from the
X-ray emission interpreted on the basis of hydrostatic equilibrium are
consistent with each other (Squires et al.\ 1996a,b).

The epoch of formation of galaxy clusters depends on cosmological
parameters, especially $\Omega_0$ (Richstone, Loeb, \& Turner 1992;
Bartelmann, Ehlers, \& Schneider 1993; Lacey \& Cole 1993, 1994).
Clusters in the local universe tend to be younger if $\Omega_0$ is
large. Such young clusters should be less relaxed and more structured
than clusters in a low density universe (Mohr et al.\ 1995; Crone,
Evrard, \& Richstone 1996). Weak lensing offers straightforward ways
to quantify cluster morphology (Wilson, Cole, \& Frenk 1996; Schneider
\& Bartelmann 1997), and therefore may be used to estimate the cosmic
density $\Omega_0$.

The dependence of cluster evolution on cosmological parameters also
has a pronounced effect on the statistics of giant arcs. Numerical
cluster simulations in different cosmological models indicate that the
observed abundance of arcs can only be reproduced in low-density
universes, $\Omega_0\sim0.3$, with vanishing cosmological constant,
$\Lambda_0\sim0$ (Bartelmann et al.\ 1997). Low-density, flat models
with $\Omega_0+\Lambda_0=1$, or Einstein-de Sitter models, produce one
or two orders of magnitude fewer arcs than observed.

\subsection{Weak Lensing by Large-Scale Structure}
\label{NB:subsect4.3}

\subsubsection{Magnification and Shear in `Empty' Fields}
\label{NB:subsubsect4.3.1}

Lensing by even larger scale structures than galaxy clusters has been
discussed in various contexts. Kristian \& Sachs (1966) and Gunn
(1967) discussed the possibility of looking for distortions in images
of background galaxies due to weak lensing by large-scale foreground
mass distributions. The idea has been revived and studied in greater
detail by Babul \& Lee (1991); Jaroszy\'nski et al.\ (1990);
Miralda-Escud\'e (1991b); Blandford et al.\ (1991); Bartelmann \&
Schneider (1991); Kaiser (1992); Seljak (1994); Villumsen (1996);
Bernardeau, van Waerbeke, \& Mellier (1997); Kaiser (1996); and Jain
\& Seljak (1997). The effect is weak---magnification and shear are
typically on the order of a few per cent---and a huge number of
galaxies has to be imaged with great care before a coherent signal can
be observed.

Despite the obvious practical difficulties, the rewards are
potentially great since the two-point correlation function of the
image distortions gives direct information on the power spectrum of
density perturbations $P(k)$ in the universe. The correlation function
of image shear, or {\em polarization\/} as it is sometimes referred to
(Blandford et al.\ 1991), has been calculated for the standard CDM
model and other popular models of the universe. Weak lensing probes
mass concentrations on large scales where the density perturbations
are still in the linear regime. Therefore, there are fewer
uncertainties in the theoretical interpretation of the phenomenon. The
problems are expected to be entirely observational.

Using a deep image of a blank field, Mould et al.\ (1994) set a limit
of $\bar p<4$ per cent for the average polarization of galaxy images
within a 4.8 arcminute field. This is consistent with most standard
models of the universe. Fahlman et al.\ (1995) claimed a tighter
bound, $\bar p<0.9$ per cent in a 2.8 arcminute field. On the other
hand, Villumsen (1995a), using the Mould et al.\ (1994) data, claimed
a detection at a level of $\bar p=(2.4\pm1.2)$ per cent ($95\%$
confidence limit). There is clearly no consensus yet, but the field is
still in its infancy.

Villumsen (1995b) has discussed how the two-point angular correlation
function of faint galaxies is changed by weak lensing and how
intrinsic clustering can be distinguished from clustering induced by
lensing. The random magnification by large-scale structures introduces
additional scatter in the magnitudes of cosmologically interesting
standard candles such as supernovae of type~Ia. For sources at
redshifts $z\sim1$, the scatter was found to be negligibly small, of
order $\Delta m\sim0.05$ magnitudes (Frieman 1996; Wambsganss et al.\
1997).

\subsubsection{Large-Scale QSO-Galaxy Correlations}
\label{NB:subsubsect4.3.2}

Fugmann (1990) noticed an excess of Lick galaxies in the vicinity of
high-redshift, radio-loud QSOs and showed that the excess reaches out
to $\sim10'$ from the QSOs. If real, this excess is most likely caused
by magnification bias due to gravitational lensing. Further, the scale
of the lens must be very large. Galaxy-sized lenses have Einstein
radii of a few arc seconds and are clearly irrelevant. The effect has
to be produced by structure on scales much larger than galaxy
clusters.

Following Fugmann's work, various other correlations of a similar
nature have been found. Bartelmann \& Schneider (1993b, 1994; see also
Bartsch, Schneider, \& Bartelmann 1997) discovered correlations
between high-redshift, radio-loud, optically bright QSOs and optical
and infrared galaxies, while Bartelmann, Schneider \& Hasinger (1994)
found correlations with diffuse X-ray emission in the $0.2-2.4$~keV
ROSAT band. Ben{\'\i}tez \& Mart{\'\i}nez-Gonz\'alez (1995, 1997)
found an excess of red galaxies from the APM catalog around radio-loud
QSOs with redshift $z\sim1$ on scales $\la10'$. Seitz \& Schneider
(1995b) found correlations between the Bartelmann \& Schneider (1993b)
sample of QSOs and foreground Zwicky clusters. They followed in part
an earlier study by Rodrigues-Williams \& Hogan (1994), who found a
highly significant correlation between optically-selected,
high-redshift QSOs and Zwicky clusters. Later, Rodrigues-Williams \&
Hawkins (1995) detected similar correlations between QSOs selected for
their optical variability and Zwicky clusters. Wu \& Han (1995)
searched for associations between distant radio-loud QSOs and
foreground Abell clusters and found a marginally significant
correlation with a subsample of QSOs.

All these results indicate that there are correlations between
background QSOs and foreground ``light'' in the optical, infrared and
soft X-ray wavebands. The angular scale of the correlations is
compatible with that expected from lensing by large-scale structures.
Bartelmann \& Schneider (1993a, see also Bartelmann 1995a for an
analytical treatment of the problem) showed that current models of
large-scale structure formation can explain the observed large-scale
QSO-galaxy associations, provided a double magnification bias
(Borgeest, von Linde, \& Refsdal 1991) is assumed. It is generally
agreed that lensing by individual clusters of galaxies is insufficient
to produce the observed effects if cluster velocity dispersions are of
order $10^3\;{\rm km\,s}^{-1}$ (e.g.\ Rodrigues-Williams \& Hogan
1994; Rodrigues-Williams \& Hawkins 1995; Wu \& Han 1995; Wu \& Fang
1996). It appears, therefore, that lensing by large-scale structures
has to be invoked to explain the observations. Bartelmann (1995a) has
shown that constraints on the density perturbation spectrum and the
bias factor of galaxy formation can be obtained from the angular
cross-correlation function between QSOs and galaxies. This calculation
was recently refined by including the non-linear growth of density
fluctuations (Sanz, Mart{\'\i}nez-Gonz\'alez, \& Ben{\'\i}tez 1997;
Dolag \& Bartelmann 1997). The non-linear effects are strong, and
provide a good fit to the observational results by Ben{\'\i}tez \&
Mart{\'\i}nez-Gonz\'alez (1995, 1997).

\subsubsection{Lensing of the Cosmic Microwave Background}
\label{NB:subsubsect4.3.3}

The random deflection of light due to large-scale structures also
affects the anisotropy of the cosmic microwave background (CMB)
radiation. The angular autocorrelation function of the CMB temperature
is only negligibly changed (Cole \& Efstathiou 1989). However,
high-order peaks in the CMB power spectrum are somewhat broadened by
lensing. This effect is weak, of order $\sim5\%$ on angular scales of
$\la10'$ (Seljak 1996; Seljak \& Zaldarriaga 1996;
Mart{\'\i}nez-Gonz\'alez, Sanz, \& Cay\'on 1997), but it could be
detected by future CMB observations, e.g.\ by the Planck Microwave
Satellite.

\subsubsection{Outlook: Detecting Dark Matter Concentrations}
\label{NB:subsubsect4.3.4}

If lensing is indeed responsible for the correlations discussed above,
other signatures of lensing should be found. Fort et al.\ (1996)
searched for shear due to weak lensing in the fields of five luminous
QSOs and found coherent signals in all five fields. In addition, they
detected foreground galaxy groups for three of the sources. Earlier,
Bonnet et al.\ (1993) had found evidence for a coherent shear pattern
in the field of the lens candidate QSO 2345$+$007. The shear was later
identified with a distant cluster (Mellier et al.\ 1994; Fischer et
al.\ 1994).

In general, it appears that looking for weak coherent image
distortions provides an excellent way of searching for otherwise
invisible dark matter concentrations. A systematic technique for this
purpose has been developed by Schneider (1996a). Weak lensing outside
cluster fields may in the near future allow observers to obtain
samples of mass concentrations which are selected purely on the basis
of their lensing effect. Such a selection would be independent of the
mass-to-light ratio, and would permit the identification and study of
nonlinear structures in the universe with unusually large
mass-to-light ratios. This would be complementary to the limits on
compact masses discussed in Sect.\ \ref{NB:subsubsect2.3.2}.

\subsubsection*{Acknowledgements}

The authors thank Rosanne Di Stefano, Andreas Huss, Chris Kochanek,
Tsafrir Kolatt, Shude Mao, Peter Schneider and Uro\v{s} Seljak for
helpful comments on the manuscript. This work was supported in part by
NSF grant AST 9423209 and by the Sonderforschungsbereich SFB 375-95 of
the Deutsche Forschungsgemeinschaft.

 %PP
%JO \end{thereferences}

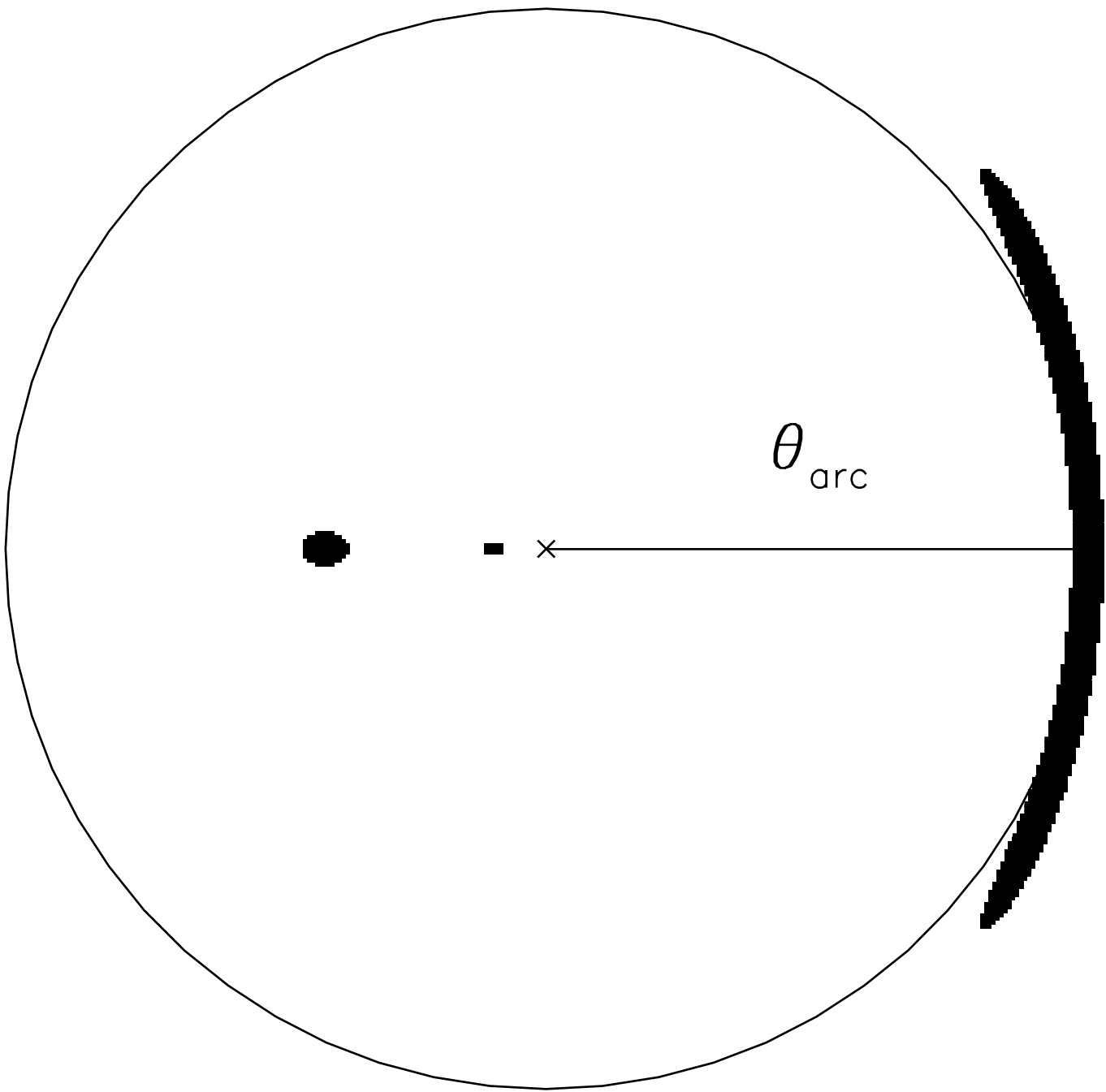
\end{document}